\documentclass{aa}
\usepackage{graphicx}
\usepackage{txfonts}
\defcitealias{dep19a}{Paper~I}
\defcitealias{dep19b}{Paper~II}
\begin{document}

\title{Plasma environment effects on K lines of astrophysical interest}

\subtitle{V. Universal formulae for ionization potential and K-threshold shifts}

\author{P. Palmeri
          \inst{1},
          J. Deprince
          \inst{1,2},
          M.~A. Bautista
           \inst{3},
          S. Fritzsche
          \inst{4,5},
          J.~A.~Garc\'ia
          \inst{2,6},
          T.~R. Kallman
          \inst{7},\\
          C. Mendoza
          \inst{3}
          \and
          P. Quinet\inst{1,8}
}

\institute{Physique Atomique et Astrophysique, Universit\'e de Mons -- UMONS, B-7000 Mons, Belgium\\
              \email{patrick.palmeri@umons.ac.be}
         \and
Cahill Center for Astronomy and Astrophysics, California Institute of Technology, Pasadena, CA 91125, USA
         \and
             Department of Physics, Western Michigan University, Kalamazoo, MI 49008, USA
         \and
             Helmholtz Institut Jena, 07743 Jena, Germany
         \and
             Theoretisch Physikalisches Institut, Friedrich Schiller Universit\"at Jena,
             07743 Jena, Germany
         \and
             Dr. Karl Remeis-Observatory and Erlangen Centre for Astroparticle Physics, Sternwartstr.~7, 96049 Bamberg, Germany
         \and
             NASA Goddard Space Flight Center, Code 662, Greenbelt, MD 20771, USA
         \and
             IPNAS, Universit\'e de Li\`ege, Sart Tilman, B-4000 Li\`ege, Belgium
}
\titlerunning{Universal formulae for ionization potential and K-threshold shifts}

\authorrunning{Deprince et al.}

\date{Received ??; accepted ??}

\abstract
   {}
   {We calculate the plasma environment effects on the ionization potentials (IPs) and K-thresholds used in the modeling of K lines for all the ions belonging to the isonuclear sequences of abundant elements apart from oxygen and iron, namely:\ carbon, silicon, calcium, chromium, and nickel. These calculations are used to extend the data points for the fits of the universal formulae, first proposed in our fourth paper of this series, to predict the IP and K-threshold lowerings in any elemental ion.}
   {We used the fully relativistic multi-configuration Dirac-Fock (MCDF) method and approximated the plasma electron-nucleus and electron-electron screenings with a time-averaged Debye-H\"uckel potential.}
   {We report the modified ionization potentials and K-threshold energies for plasmas characterized by electron temperatures and densities in the ranges of $10^5{-}10^7$~K and $10^{18}{-}10^{22}$~cm$^{-3}$. In addition, the improved universal fitting formulae are obtained.}
   {We conclude that since explicit calculations of the atomic structures for each ion of each element under different plasma conditions is impractical, the use of these universal formulae for predicting the IP and K-threshold lowerings in plasma modeling codes is still recommended. However, their comparatively moderate to low accuracies may affect the predicted opacities with regard to certain cases under extreme plasma conditions that are characterized by a plasma screening parameter of $\mu > 0.2$~a.u., especially for the K-thresholds. }


\maketitle
%

\section{Introduction}

Supersolar abundances have been inferred from K lines of different elements observed in the X-ray spectra of X-ray binaries (XRB) and active galactic nuclei (AGN) (see e.g., \citet{kal09,don20,wal20,fuk21}). These absorption and emission features can occur in the inner regions of the black-hole accretion disks where the plasma densities are predicted to range from $10^{15}$ to $10^{22}$~cm$^{-3}$ \citep{sch13}. The emerging photons can be recorded with current space observatories such as {\it XMM-Newton}, {\it NuSTAR}, and {\it Chandra}. In addition, synthetic spectra can provide measures of the composition, temperature, and degree of ionization of the plasma \citep{ros05, gar10}.  Nevertheless, the great majority of the atomic parameters used for spectral modeling that involves K-shell processes do not take density effects into account and, therefore, their usefulness is comprised in abundance determinations beyond densities of $10^{18}$~cm$^{-3}$ \citep{smi14}.

In a series of papers dedicated to plasma density effects on the atomic parameters used to model K lines in ions of astrophysical interest, the ionization potentials (IPs), K-thresholds, transition wavelengths, radiative emission rates, and Auger widths have been computed with the relativistic multiconfiguration Dirac--Fock (MCDF) method \citep{gra80,mck80,gra88}, as implemented in the GRASP92 \citep{par96} and RATIP \citep{fri12} atomic structure packages. The plasma electron--nucleus and electron--electron screenings are approximated with a time-averaged Debye-H\"uckel (DH) potential. The datasets comprise the following ionic species:
\ion{O}{i} -- \ion{O}{vii}, by \citet[][hereafter Paper~I]{dep19a};
\ion{Fe}{xvii} -- \ion{Fe}{xxv}, by \citet[][hereafter Paper~II]{dep19b};
\ion{Fe}{ix} -- \ion{Fe}{xvi}, by \citet[][hereafter Paper~III]{dep20a};
\ion{and Fe}{ii} -- \ion{Fe}{viii}, by \citet[][hereafter Paper~IV]{dep20b}.

In this fifth paper of the series, the universal fitting formulae for ionization potential (IP) and K-threshold shifts proposed in Paper~IV are improved thanks to subsequent MCDF/RATIP computations of these above-mentioned parameters in other representative cosmically abundant elements, namely:\ carbon ($Z=6$), silicon ($Z=14$), calcium ($Z=20$), chromium ($Z=24$), and nickel ($Z=28$). These plasma effects are expected to be the main ones with respect to the potential alteration of the ionization balance and opacities.

\section{Theoretical method used}

In this section, we outline the main changes introduced in the MCDF formalism to take into account the density effects in a weakly coupled plasma. Papers~I--II and IV contain more details on this, along with tests of the validity of the method.

The Debye-H\"uckel (DH) screened Dirac--Coulomb Hamiltonian \citep{sah06} is expressed as:
\begin{equation}
H^{DH}_{DC}=\sum_i c \vec{\alpha_i} \cdot \vec{p_i}+ \beta_i c^2 - \frac{Z}{r_i}
e^{-\mu r_i}
+ \sum_{i>j} \frac{1}{r_{ij}} e^{-\mu r_{ij}}\ ,
\label{dh}
\end{equation}
where $r_{ij}=|\vec{r}_i-\vec{r}_j|$ and the plasma screening parameter $\mu$ is the inverse of the Debye shielding length $\lambda_D$. The screening parameter can be given in atomic units (a.u.) in terms of the plasma electron density, $n_e$, and temperature, $T_e$, as
\begin{equation}
\mu = \frac{1}{\lambda_D} = \sqrt{\frac{4\pi n_e}{k T_e}}\ .
\label{screen}
\end{equation}
For the typical plasma conditions in black-hole accretion disks, $T_e\sim  10^5{-}10^7$~K and $n_e\sim 10^{18}{-}10^{22}$~cm$^{-3}$ \citep{sch13}, the screening parameter $\mu$ falls in the interval: $0.0 \leq \mu \leq 0.24$~a.u. The DH screening theory is only valid for a weakly coupled plasma where its thermal energy dominates its electrostatic energy. This can be parameterized by the plasma coupling parameter $\Gamma,$ defined as:
\begin{equation}
\Gamma = \frac{Z^{*2} e^2}{4\pi\epsilon_0 d kT_e}
,\end{equation}
where $Z^*$ is the average plasma ionic charge or ionization:
\begin{equation}
Z^*=\frac{\sum_{i,X} z_{i,X} n_{i,X}}{\sum_{i,X} n_{i,X}}
\label{z*}
,\end{equation}
with $n_{i,X}$ as the number density of an ion, $i,$ of an element, $X,$ bearing a positive charge, $z_{i,X}$,
and $d$ as a measurement of the interparticle distance:
\begin{equation}
d=\left(\frac{3}{4\pi \sum_{i,X} n_{i,X}}\right)^{1/3}~.
\label{d}
\end{equation}
The plasma neutrality implies:
\begin{equation}
\sum_{i,X} z_{i,X} n_{i,X} = n_e
,\end{equation}
and therefore Eqs. (\ref{z*}) and (\ref{d}) can be rewritten as:
\begin{equation}
Z^*=\frac{n_e}{\sum_{i,X} n_{i,X}}
,\end{equation}
\begin{equation}
d=\left(\frac{3Z^*}{4\pi n_e}\right)^{1/3}
.\end{equation}
For a fully ionized hydrogen plasma with $Z^* = 1$, the plasma coupling parameter, considering typical conditions in accretion disks, is well below 1 falling in the interval $0.0003\leq \Gamma \leq 0.3$ which is characteristic of a weakly coupled plasma. Furthermore, for a more realistic cosmological plasma with a mixture of 90\% hydrogen and 10\% of helium by number (both fully ionized), the average plasma ionization would be $Z^*~=~0.9~+~2~\times~0.1~=~1.1$. Hence, the above interval limits on the plasma coupling parameter would have to be multiplied by a factor $Z^{*2/3} \sim 1.07$ and would still agree with a weakly coupled plasma (see our Paper~IV for further details).

In the present study, the active space (AS) method was used in order to obtain the MCDF multiconfigurational expansions for the all the ionization stages of the isonuclear sequences of C, Si, Ca, Cr, and Ni, analogously to those of O and Fe in the recent past (see Papers I--IV). This method consists of exciting the electrons from the reference configurations to a given active set of orbitals. For almost all the ions studied in this work, the AS was built by considering all the single and double excitations from some reference configurations (always including the ground ones) up to the $n=3$ orbitals (when only K- and L-shell electrons are involved in the ground configuration) or up to the $n=3$ and 4s orbitals (when K-, L-, and M-shell electrons are involved in the ground configuration). The reference configurations were taken as the ground one along with those obtained by considering single electron excitations leaving a single hole in the 1s, 2p, and 3p orbitals (when involved in the ground configuration). In some cases, namely for the ions close to the neutral end (i.e., for \ion{Ni}{i} -- \ion{Ni}{iv}, \ion{Cr}{i} -- \ion{Cr}{ii}, \ion{Ca}{i} -- \ion{Ca}{ii}, \ion{Si}{i} -- \ion{Si}{ii}), only single excitations from the reference configurations were considered. In Table~\ref{as}, we report for each ion grouped by number of electrons, $N$, the reference configurations, the active orbital sets and the final numbers of configuration state functions (CSF),$\Phi $, generated to build the atomic state functions (ASF), $\Psi$ , in the MCDF expansions as:
\begin{equation}
\Psi = \sum^{N_{CSF}}_{k = 1}~c_k~\Phi_k \label{mcdf}
,\end{equation}
where $c_k$ are the mixing coefficients. 

The computations were carried out with the extended average level (EAL) option, optimizing a weighted trace of the Hamiltonian using level weights proportional to $2J+1$, and they were completed with the inclusion of the relativistic two-body Breit interaction and the quantum electrodynamic corrections (QED) due to self-energy and vacuum polarization. The MCDF ionic bound states generated by the GRASP92 code were then used in the RATIP program to compute the atomic structures diagonalizing the DH screened Dirac--Coulomb Hamiltonian (Eq.~(\ref{dh})), from which were then obtained the ionization potential and the K-shell threshold energies. Plasma environment effects are computed for a Debye screening parameter in the range $0 \leq \mu \leq 0.25$ a.u.

\begin{table*}[!ht]
  \caption{Reference configurations and active orbital sets used to build up the MCDF active space by single and double electron excitations to the corresponding active orbital sets along with the number of configuration state functions (CSFs), $N_{CSF}$, generated for the MCDF expansions in carbon, silicon, calcium, chromium, and nickel ions with $N = 2-28$ electrons. \label{as}}
  \centering
  \small
  \begin{tabular}{rlllr}
  \hline\hline
  \noalign{\smallskip}
  $N$ & Ions & Reference configurations$^a$ & Active orbital set & $N_{CSF}$\\
  \hline
  \noalign{\smallskip}
        2 & \ion{C}{v}, \ion{Si}{xiii}, \ion{Ca}{xix}, \ion{Cr}{xxiii}, 
        \ion{Ni}{xxvii}         
        & ${\rm 1s^2}$,
        ${\rm 1s2s}$,
        ${\rm 1s2p}$
                & ${\rm \{1s,2s,2p,3s,3p,3d\}}$
                & 98 \\        
        3 & \ion{C}{iv}, \ion{Si}{xii}, \ion{Ca}{xviii}, \ion{Cr}{xxii}, 
        \ion{Ni}{xxvi}    
        & ${\rm 1s^22s}$,
        ${\rm 1s^22p}$,
        ${\rm 1s2s^2}$,
        ${\rm 1s2s2p}$,
        ${\rm 1s2p^2}$
                & ${\rm \{1s,2s,2p,3s,3p,3d\}}$
                & 515\\
        4 & \ion{C}{iii}, \ion{Si}{xi}, \ion{Ca}{xvii}, \ion{Cr}{xxi}, 
        \ion{Ni}{xxv} 
        & ${\rm 1s^22s^2}$,
        ${\rm 1s^22s2p}$,
        ${\rm 1s^22p^2}$,
        ${\rm 1s2s^22p}$,
        ${\rm 1s2s2p^2}$,
        ${\rm 1s2p^3}$
                & ${\rm \{1s,2s,2p,3s,3p,3d\}}$
                & 1847\\
        5 & \ion{C}{ii}, \ion{Si}{x}, \ion{Ca}{xvi}, \ion{Cr}{xx}, 
        \ion{Ni}{xxiv} 
        & ${\rm 1s^22s^22p}$,
        ${\rm 1s^22s2p^2}$,
        ${\rm 1s^22p^3}$,
        ${\rm 1s2s^22p^2}$,
        ${\rm 1s2s2p^3}$,
        ${\rm 1s2p^4}$
                & ${\rm \{1s,2s,2p,3s,3p,3d\}}$
                & 4107\\
        6 & \ion{C}{i}, \ion{Si}{ix}, \ion{Ca}{xv}, \ion{Cr}{xix}, 
        \ion{Ni}{xxiii}  
        & ${\rm 1s^22s^22p^2}$,             
        ${\rm 1s^22s2p^3}$,
        ${\rm 1s^22p^4}$,
        ${\rm 1s2s^22p^3}$,
        ${\rm 1s2s2p^4}$,
        ${\rm 1s2p^5}$
                & ${\rm \{1s,2s,2p,3s,3p,3d\}}$
                & 6730\\
        7 & \ion{Si}{viii}, \ion{Ca}{xiv}, \ion{Cr}{xviii}, 
        \ion{Ni}{xxii} 
        & ${\rm 1s^22s^22p^3}$,             
        ${\rm 1s^22s2p^4}$,
        ${\rm 1s^22p^5}$,
        ${\rm 1s2s^22p^4}$,
        ${\rm 1s2s2p^5}$,
        ${\rm 1s2p^6}$
                & ${\rm \{1s,2s,2p,3s,3p,3d\}}$
                & 7389\\
        8 & \ion{Si}{vii}, \ion{Ca}{xiii}, \ion{Cr}{xvii}, 
        \ion{Ni}{xxi} 
        & ${\rm 1s^22s^22p^4}$,             
        ${\rm 1s^22s2p^5}$,
        ${\rm 1s^22p^6}$,
        ${\rm 1s2s^22p^5}$,
        ${\rm 1s2s2p^6}$
                & ${\rm \{1s,2s,2p,3s,3p,3d\}}$
                & 6013\\
        9 & \ion{Si}{vi}, \ion{Ca}{xii}, \ion{Cr}{xvi}, 
        \ion{Ni}{xx} 
        & ${\rm 1s^22s^22p^5}$,             
        ${\rm 1s^22s2p^6}$,
        ${\rm 1s2s^22p^6}$
                & ${\rm \{1s,2s,2p,3s,3p,3d\}}$
                & 2638\\
        10 & \ion{Si}{v}, \ion{Ca}{xi}, \ion{Cr}{xv}, 
        \ion{Ni}{xix} 
        & ${\rm 1s^22s^22p^6}$,             
        ${\rm 1s^22s^22p^53s}$,
        ${\rm 1s^22s^22p^53p}$,
        ${\rm 1s2s^22p^63s}$,
        ${\rm 1s2s^22p^63p}$
                & ${\rm \{1s,2s,2p,3s,3p,3d\}}$
                & 12564\\
        11 & \ion{Si}{iv}, \ion{Ca}{x}, \ion{Cr}{xiv}, 
        \ion{Ni}{xviii} 
        & ${\rm 3s}$,
                ${\rm [2p]3s^2}$,
                ${\rm [2p]3s3p}$,
                ${\rm [1s]3s^2}$,
                ${\rm [1s]3s3p}$
                & ${\rm \{1s,2s,2p,3s,3p,3d,4s\}}$
                & 25914  \\
                12 & \ion{Si}{iii}, \ion{Ca}{ix}, \ion{Cr}{xiii}, 
        \ion{Ni}{xvii} 
                & ${\rm 3s^2}$,
                ${\rm [2p]3s^23p}$,
                ${\rm [1s]3s^23p}$
            & ${\rm \{1s,2s,2p,3s,3p,3d,4s\}}$
                & 16853  \\
                13 & \ion{Si}{ii} 
                & ${\rm 3p}$,
                ${\rm [2p]3p^2}$,
                ${\rm [1s]3p^2}$
            & ${\rm \{1s,2s,2p,3s,3p,3d,4s\}}$
                & 35109$^*$  \\
                13 & \ion{Ca}{viii}, \ion{Cr}{xii}, 
        \ion{Ni}{xvi} 
                & ${\rm 3p}$,
                ${\rm [2p]3p^2}$,
                ${\rm [1s]3p^2}$
            & ${\rm \{1s,2s,2p,3s,3p,3d,4s\}}$
                & 35109  \\
                14 & \ion{Si}{i} 
                & ${\rm 3p^2}$,
                ${\rm [2p]3p^3}$,
                ${\rm [1s]3p^3}$
                & ${\rm \{1s,2s,2p,3s,3p,3d,4s\}}$
                & 46771$^*$  \\
                14 & \ion{Ca}{vii}, \ion{Cr}{xi}, 
        \ion{Ni}{xv} 
                & ${\rm 3p^2}$,
                ${\rm [2p]3p^3}$,
                ${\rm [1s]3p^3}$
                & ${\rm \{1s,2s,2p,3s,3p,3d,4s\}}$
                & 46771  \\
                15 & \ion{Ca}{vi}, \ion{Cr}{x}, 
        \ion{Ni}{xiv} 
                & ${\rm 3p^3}$,
            ${\rm [2p]3p^4}$,
            ${\rm [1s]3p^4}$
            & ${\rm \{1s,2s,2p,3s,3p,3d,4s\}}$
                & 37967  \\
                16 & \ion{Ca}{v}, \ion{Cr}{ix}, 
        \ion{Ni}{xiii} 
                & ${\rm 3p^4}$,
                ${\rm [2p]3p^5}$,
                ${\rm [1s]3p^5}$
            & ${\rm \{1s,2s,2p,3s,3p,3d,4s\}}$
                & 12981  \\
                17 & \ion{Ca}{iv}, \ion{Cr}{viii}, 
        \ion{Ni}{xii} 
                & ${\rm 3p^5}$,
                ${\rm [2p]3p^6}$,
                ${\rm [1s]3p^6}$                
                & ${\rm \{1s,2s,2p,3s,3p,3d,4s\}}$
                & 6312  \\
                18 & \ion{Ca}{iii}, \ion{Cr}{vii}, 
        \ion{Ni}{xi} 
                & ${\rm 3p^6}$,
                ${\rm [3p]3d}$,
                ${\rm [2p]3d}$,
                ${\rm [1s]3d}$
                & ${\rm \{1s,2s,2p,3s,3p,3d,4s\}}$
                & 20009  \\
                19 &    \ion{Ca}{ii}  
                & ${\rm 4s}$,
                        ${\rm [3p]4s^2}$,
                        ${\rm [2p]4s^2}$,
                        ${\rm [1s]4s^2}$
                & ${\rm \{1s,2s,2p,3s,3p,3d,4s\}}$      
                        &  43271$^*$ \\
                19 &    \ion{Cr}{vi}, \ion{Ni}{x}  
                & ${\rm 3d}$,
                        ${\rm [3p]3d^2}$,
                        ${\rm [2p]3d^2}$,
                        ${\rm [1s]3d^2}$
                & ${\rm \{1s,2s,2p,3s,3p,3d,4s\}}$      
                        &  43271 \\
                20 &    \ion{Ca}{i} 
                & ${\rm 4s^2}$,
                        ${\rm [3p]3d4s^2}$,
                        ${\rm [2p]3d4s^2}$,
                        ${\rm [1s]3d4s^2}$
                & ${\rm \{1s,2s,2p,3s,3p,3d,4s\}}$
                        &  71135$^*$ \\
                20 &    \ion{Cr}{v}, \ion{Ni}{ix} 
                & ${\rm 3d^2}$,
                        ${\rm [3p]3d^3}$,
                        ${\rm [2p]3d^3}$,
                        ${\rm [1s]3d^3}$
                & ${\rm \{1s,2s,2p,3s,3p,3d,4s\}}$
                        &  71135 \\
                21 &    \ion{Cr}{iv}, \ion{Ni}{viii} 
                & ${\rm 3d^3}$,
                        ${\rm [3p]3d^4}$,
                        ${\rm [2p]3d^4}$,
                        ${\rm [1s]3d^4}$
                & ${\rm \{1s,2s,2p,3s,3p,3d,4s\}}$
                        &  85798 \\
                22 &    \ion{Cr}{iii}, \ion{Ni}{vii} 
                & ${\rm 3d^4}$,
                        ${\rm [3p]3d^5}$,
                        ${\rm [2p]3d^5}$,
                        ${\rm [1s]3d^5}$
                & ${\rm \{1s,2s,2p,3s,3p,3d,4s\}}$
                        & 81237  \\
                23 &    \ion{Cr}{ii} 
                & ${\rm 3d^5}$,
                        ${\rm [3p]3d^6}$,
                        ${\rm [2p]3d^6}$,
                        ${\rm [1s]3d^6}$
                & ${\rm \{1s,2s,2p,3s,3p,3d,4s\}}$
                        & 2681$^*$  \\
                23 &    \ion{Ni}{vi} 
                & ${\rm 3d^5}$,
                        ${\rm [3p]3d^6}$,
                        ${\rm [2p]3d^6}$,
                        ${\rm [1s]3d^6}$
                & ${\rm \{1s,2s,2p,3s,3p,3d,4s\}}$
                        & 57189  \\
                24 &    \ion{Cr}{i} 
                & ${\rm 3d^54s}$,
                        ${\rm [3p]3d^64s}$,
                        ${\rm [2p]3d^64s}$,
                        ${\rm [1s]3d^64s}$
                & ${\rm \{1s,2s,2p,3s,3p,3d,4s\}}$                      
                         &8660$^*$ \\
                24 &    \ion{Ni}{v} 
                & ${\rm 3d^6}$,
                        ${\rm [3p]3d^7}$,
                        ${\rm [2p]3d^7}$,
                        ${\rm [1s]3d^7}$
                & ${\rm \{1s,2s,2p,3s,3p,3d,4s\}}$                      
                         &8660 \\
                25 &    \ion{Ni}{iv} 
                & ${\rm 3d^7}$,
                        ${\rm [3p]3d^8}$,
                        ${\rm [2p]3d^8}$,
                        ${\rm [1s]3d^8}$ 
                & ${\rm \{1s,2s,2p,3s,3p,3d,4s\}}$                      
                        & 2200$^*$\\
                26 &    \ion{Ni}{iii} 
                & ${\rm 3d^8}$,
                        ${\rm [3p]3d^9}$,
                         ${\rm [2p]3d^9}$,
                         ${\rm [1s]3d^9}$
                & ${\rm \{1s,2s,2p,3s,3p,3d,4s\}}$                      
                        & 700$^*$\\
                27 &    \ion{Ni}{ii} 
                & ${\rm 3d^9}$,
                        ${\rm [3p]3d^{10}}$,
                         ${\rm [2p]3d^{10}}$,
                         ${\rm [1s]3d^{10}}$
                & ${\rm \{1s,2s,2p,3s,3p,3d,4s\}}$                      
                        & 144$^*$\\
                28 &    \ion{Ni}{i} 
                & ${\rm 3d^84s^2}$,
                        ${\rm [3p]3d^94s^2}$,
                         ${\rm [2p]3d^94s^2}$,
                         ${\rm [1s]3d^94s^2}$
                & ${\rm \{1s,2s,2p,3s,3p,3d,4s\}}$                      
                        & 88$^*$\\
  \hline
  \end{tabular}
        \tablefoot{\tablefoottext{a}{[$n\ell$] means a hole in the $n\ell$ subshell.}
        \tablefoottext{*}{Only single excitations were considered.}}
\end{table*}

\section{Results and discussion}

\subsection{Ionization potentials and IP shifts}

In Tables~\ref{IPC}--\ref{IPNi}, the MCDF/RATIP ionization potentials (IPs) are reported for plasma screening parameters $\mu=0.0$ a.u., that is, in the isolated atom case, $\mu =0.1$ a.u., and $\mu = 0.25$ a.u., along with the corresponding values recommended by the National Institute of Standards and Technology \citep[][NIST]{nist}, respectively, for carbon, silicon, calcium, chromium, and nickel. The differences between our theoretical values for $\mu=0$~a.u. and the NIST IPs are on average within $\sim \pm$3~eV. In terms of relative differences, they can span from less than a few \% for the highly-charged ions, that is, with $Z_{\rm eff}= Z-N+1 > 10$ where $Z_{\rm eff}$ is the effective charge and $N$ is the number of bound electrons, up to 22 \% in \ion{Ni}{i} and 38\% in \ion{Cr}{i}. The latter is the result of numerical difficulties in computing the complex atomic structures involving half-filled 3d subshell and in neutral ends of isonuclear sequences in iron group elements such as chromium and nickel, where the three lower even-parity configurations ${\rm 3d^w}$, ${\rm 3d^{w-1}4s,}$ and ${\rm 3d^{w-2}4s^2}$ overlap and are highly mixed. 

\begin{figure}[!ht]
  \centering
  \includegraphics*[pagebox=mediabox, width=\columnwidth]{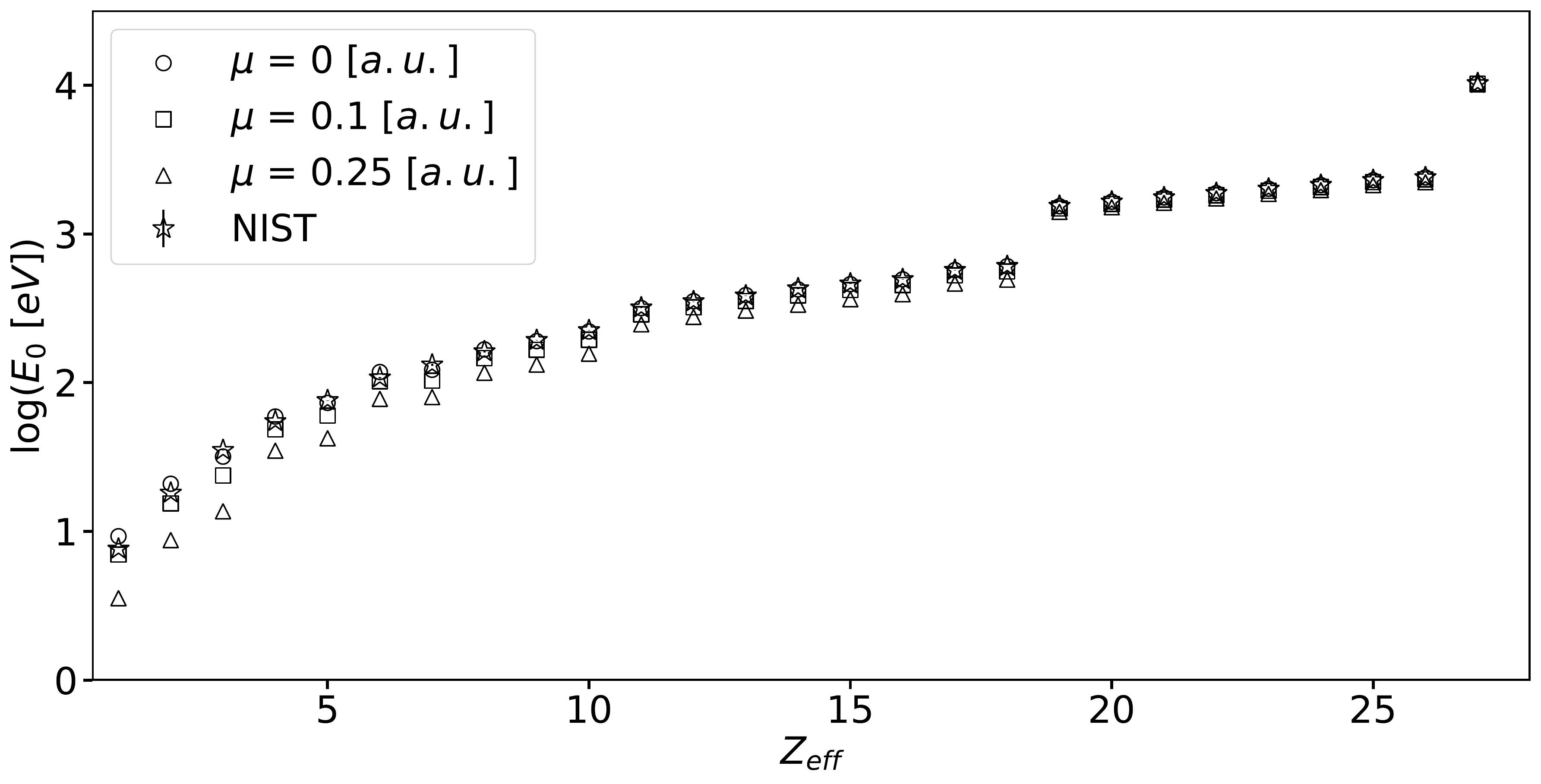}
  \caption{Ionization potential, $E_0$ in eV, in logarithmic scale as a function of the effective charge, $Z_{\rm eff}$, in ions belonging to the nickel isonuclear sequence. The NIST error bars are too small to appear in the figure.} \label{figIPvsZeff_nickel}
\end{figure}

Figure~\ref{figIPvsZeff_nickel} shows the IP, $E_0$ in eV, in logarithmic scale as function of $Z_{\rm eff}$, for ions belonging to nickel isonuclear sequence chosen as an example. The MCDF/RATIP IPs calculated in this work for $\mu = 0$, 0.1 and 0.25~a.u. are plotted along with the corresponding recommended values of NIST \citep{nist}. The latter are to be compared with our calculations in the isolated atom case, namely, for $\mu = 0$~a.u. The NIST error bars are too small to be seen in the figure.
We can observe that all four curves follow the same trend with the curves for $\mu=0.1$~a.u. and $\mu=0.25$~a.u., systematically downshifted with respect to both the NIST and the isolated-atom case ones. Also, two big jumps are seen between $Z_{\rm eff}= 26$ and $Z_{\rm eff}= 27$ and between $Z_{\rm eff}= 18$ and $Z_{\rm eff}= 19$. These correspond to respectively the transition from the closure of the K-shell and the starting of the filling of the L-shell, and the transition from the closure of the L-shell and the starting of the filling of the M-shell. These features are similar to the ones observed in the oxygen (Paper~I) and iron isonuclear sequences (Papers~II-IV).

\begin{figure}[!ht]
  \centering
  \includegraphics*[pagebox=mediabox, width=\columnwidth]{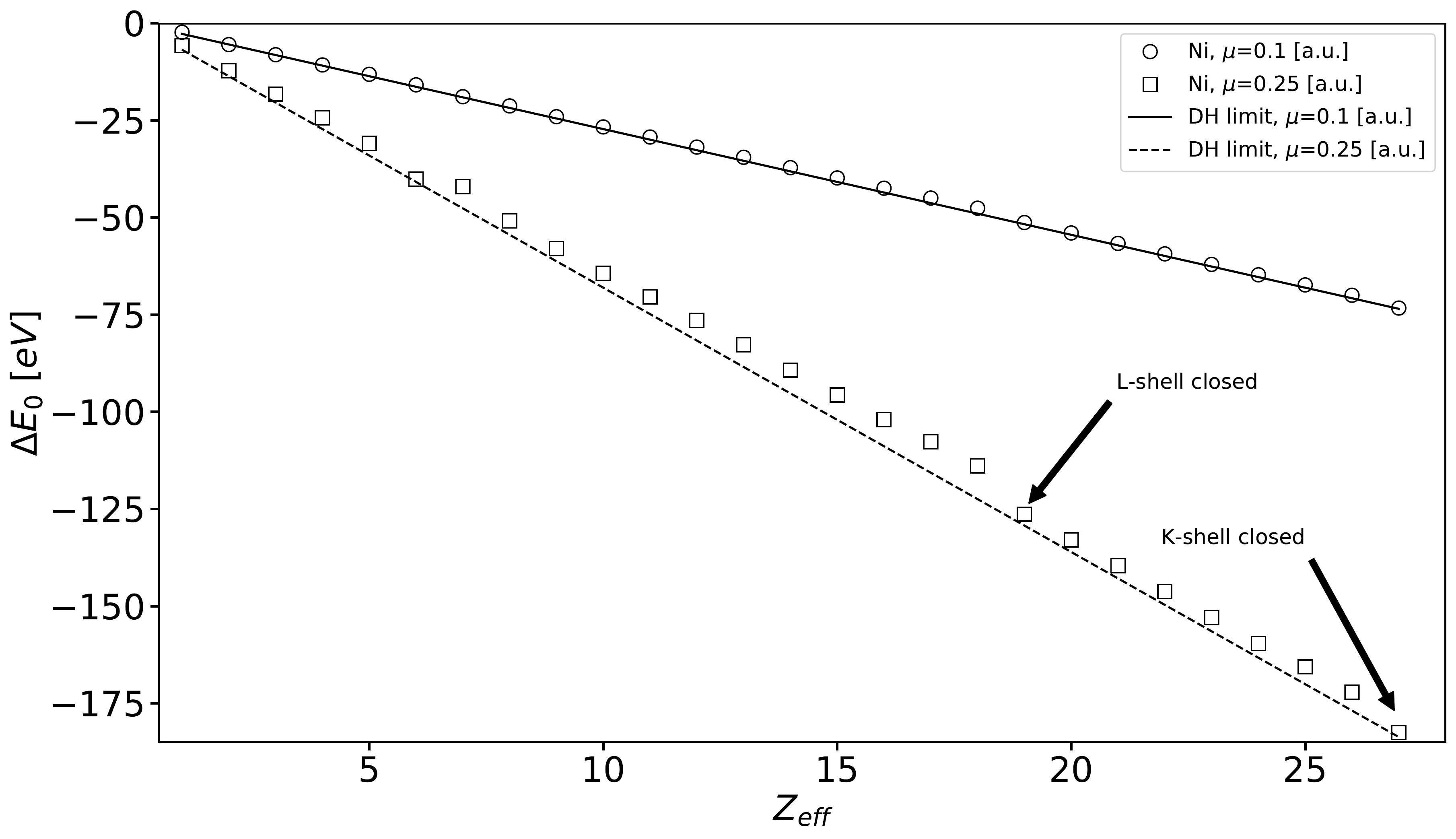}
  \caption{Ionization potential shift, $\Delta E_0$ in eV, as a function of the effective charge, $Z_{\rm eff}$, in ions belonging to the nickel isonuclear sequence.  Circles: MCDF/RATIP method for $\mu = 0.1$~a.u. Squares: MCDF/RATIP method for $\mu = 0.25$~a.u. Solid line: Debye-H\"uckel limit for $\mu = 0.1$~a.u. Dashed line: Debye-H\"uckel limit for $\mu = 0.25$~a.u.} \label{figIPDvsZeff_nickel}
\end{figure}

\begin{figure}[!ht]
  \centering
  \includegraphics*[pagebox=mediabox, width=\columnwidth]{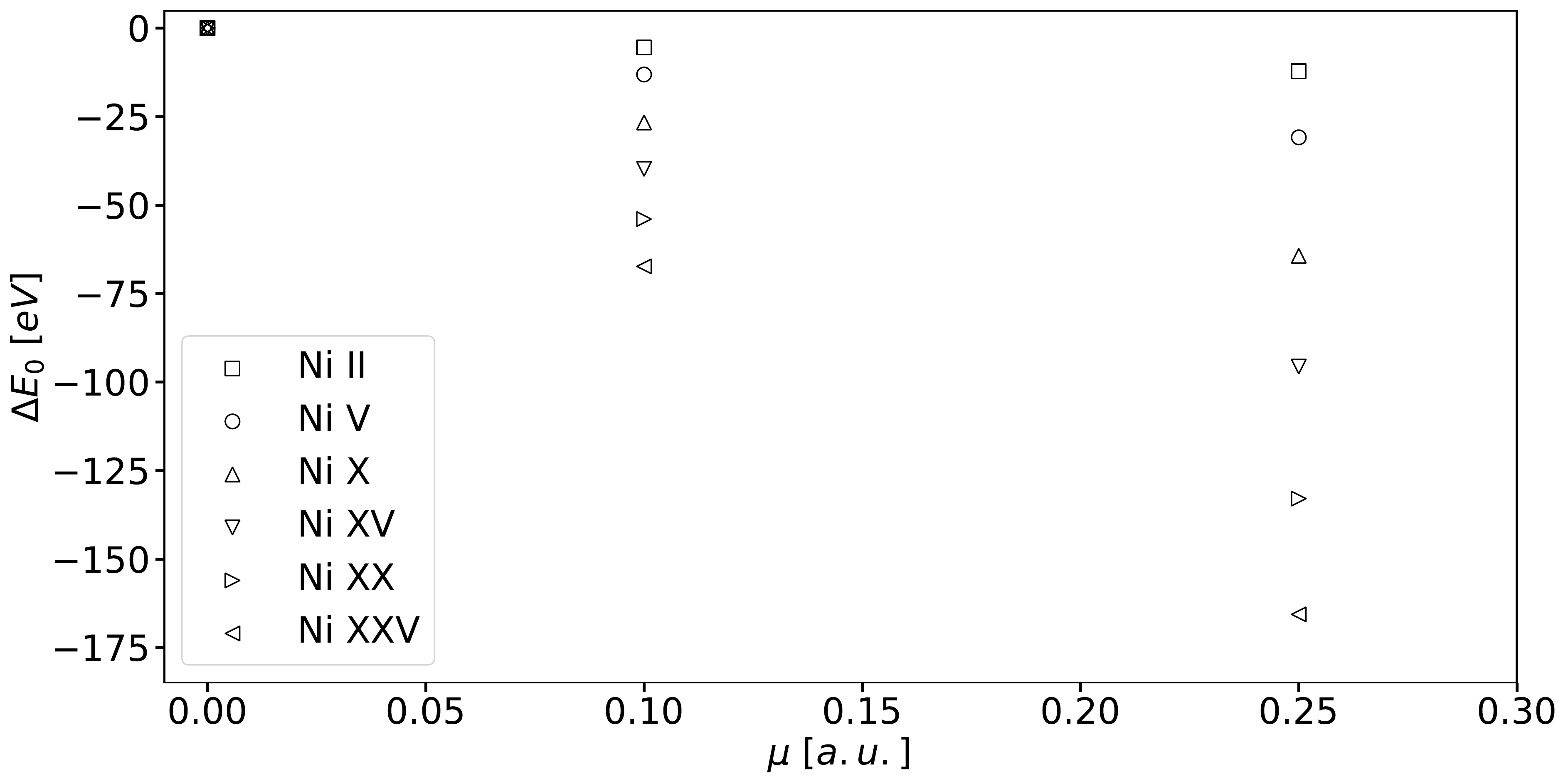}
  \caption{MCDF/RATIP ionization potential shift, $\Delta E_0$ in eV, as a function of the plasma screening parameter $\mu$ in a.u., in some ions belonging to the nickel isonuclear sequence.} \label{figIPDvsmu_nickel}
\end{figure}

In Figure~\ref{figIPDvsZeff_nickel}, the MCDF/RATIP IP shift, $\Delta E_0 = E_0(\mu) - E_0(\mu = 0)$ in eV, with respect to the isolated atom plasma condition ($\mu = 0$~a.u.) is plotted as function of the effective charge in the same example (\ion{Ni}{i}--\ion{Ni}{xxvii}) for plasma screen parameters of 0.1~a.u. and 0.25~a.u. It can be seen that this quantity varies linearly with $Z_{\rm eff}$ with steeper slope for larger $\mu$. Moreover,
it has also a linear trend with the plasma screening parameter as displayed for the nickel ions in Figure~\ref{figIPDvsmu_nickel}. This is in agreement with the corresponding Debye-H\"uckel (DH) limit \citep{sp66,crow14}. This is also shown in Figure~\ref{figIPDvsZeff_nickel} as solid and dashed lines as given below:
\begin{equation}
\Delta E_0^{DH} =~-27.2116~\mu~Z_{\rm eff} \ ,
\label{DHlimitIP}
\end{equation}
where $\Delta E_0^{DH}$ is in eV and $\mu$ in a.u. Besides, closed-shell effects are clearly marked in Figure~\ref{figIPDvsZeff_nickel} for $\mu= 0.25$~a.u. with small departures from the DH limit at $Z_{\rm eff} = 18$ and 26. 

\subsection{K-threshold energies and shifts}

\begin{figure}[!ht]
  \centering
  \includegraphics*[pagebox=mediabox, width=\columnwidth]{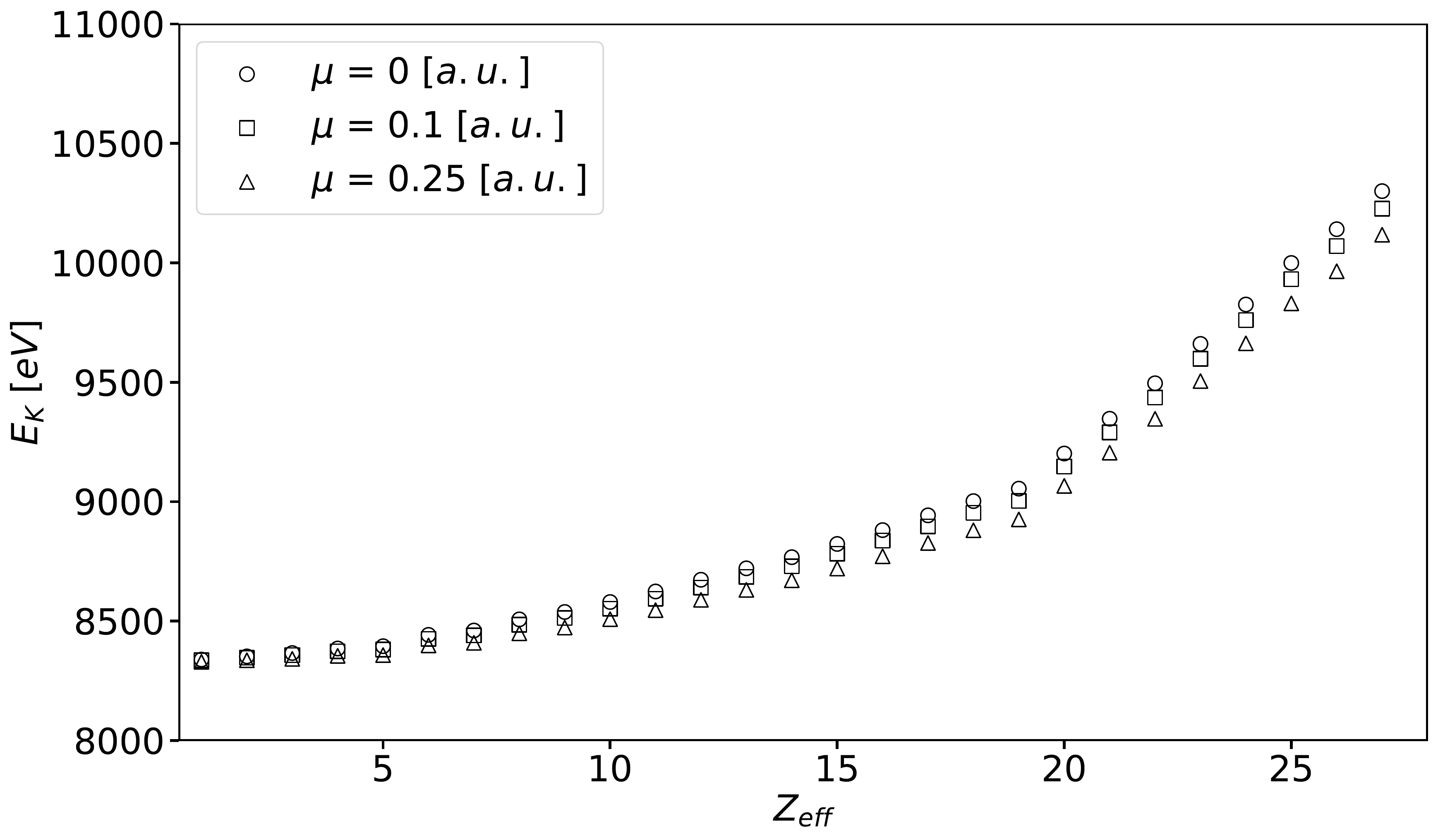}
  \caption{MCDF/RATIP K-threshold energy, $E_K$ in eV, as a function of the effective charge, $Z_{\rm eff}$, in ions belonging to the nickel isonuclear sequence.} \label{figKTvsZeff_nickel}
\end{figure}

\begin{figure}[!ht]
  \centering
  \includegraphics*[pagebox=mediabox, width=\columnwidth]{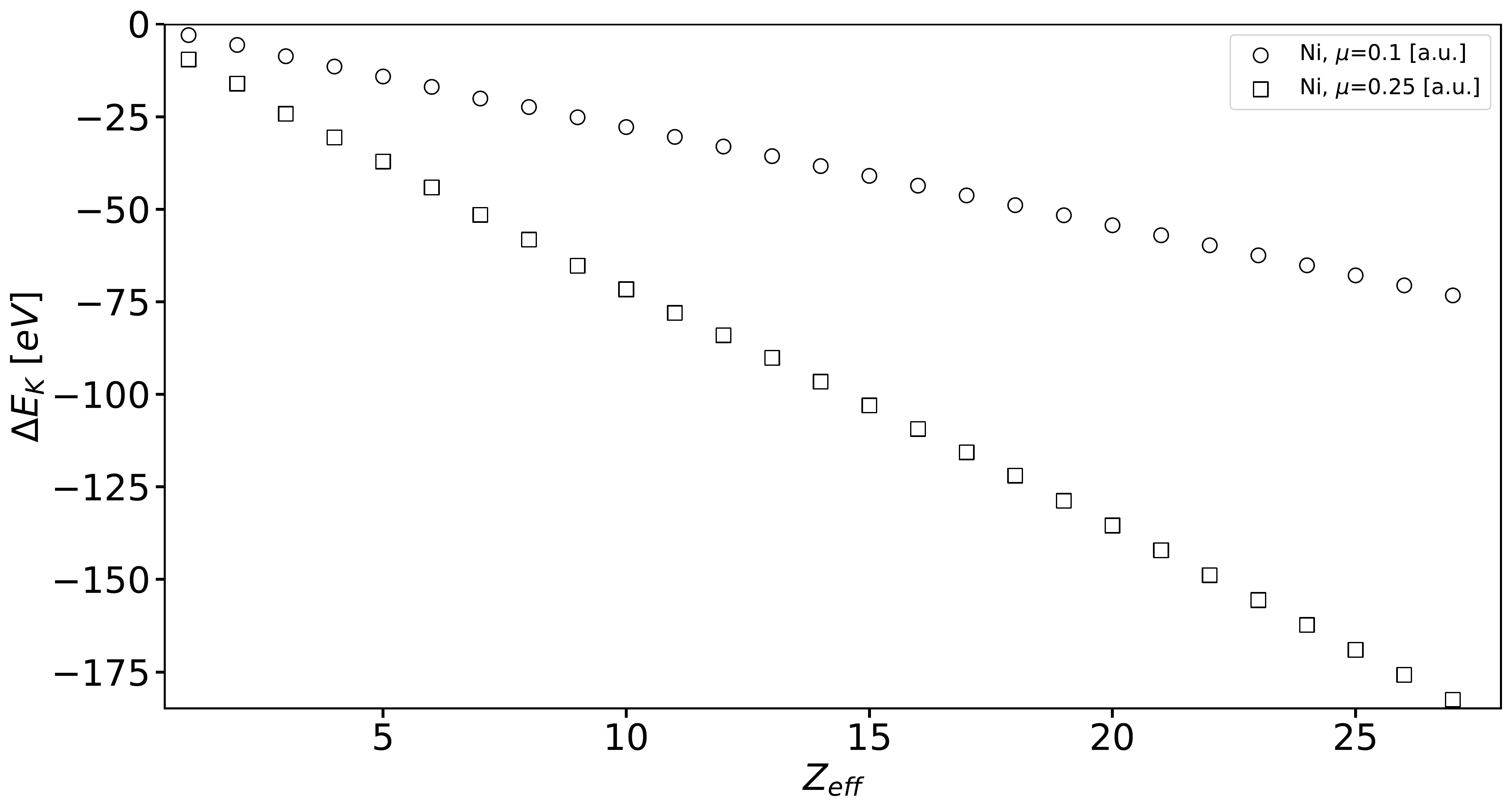}
  \caption{MCDF/RATIP K-threshold energy shift, $\Delta E_K$ in eV, as a function of the effective charge, $Z_{\rm eff}$, in ions belonging to the nickel isonuclear sequence.} \label{figKTDvsZeff_nickel}
\end{figure}

\begin{figure}[!ht]
  \centering
  \includegraphics*[pagebox=mediabox, width=\columnwidth]{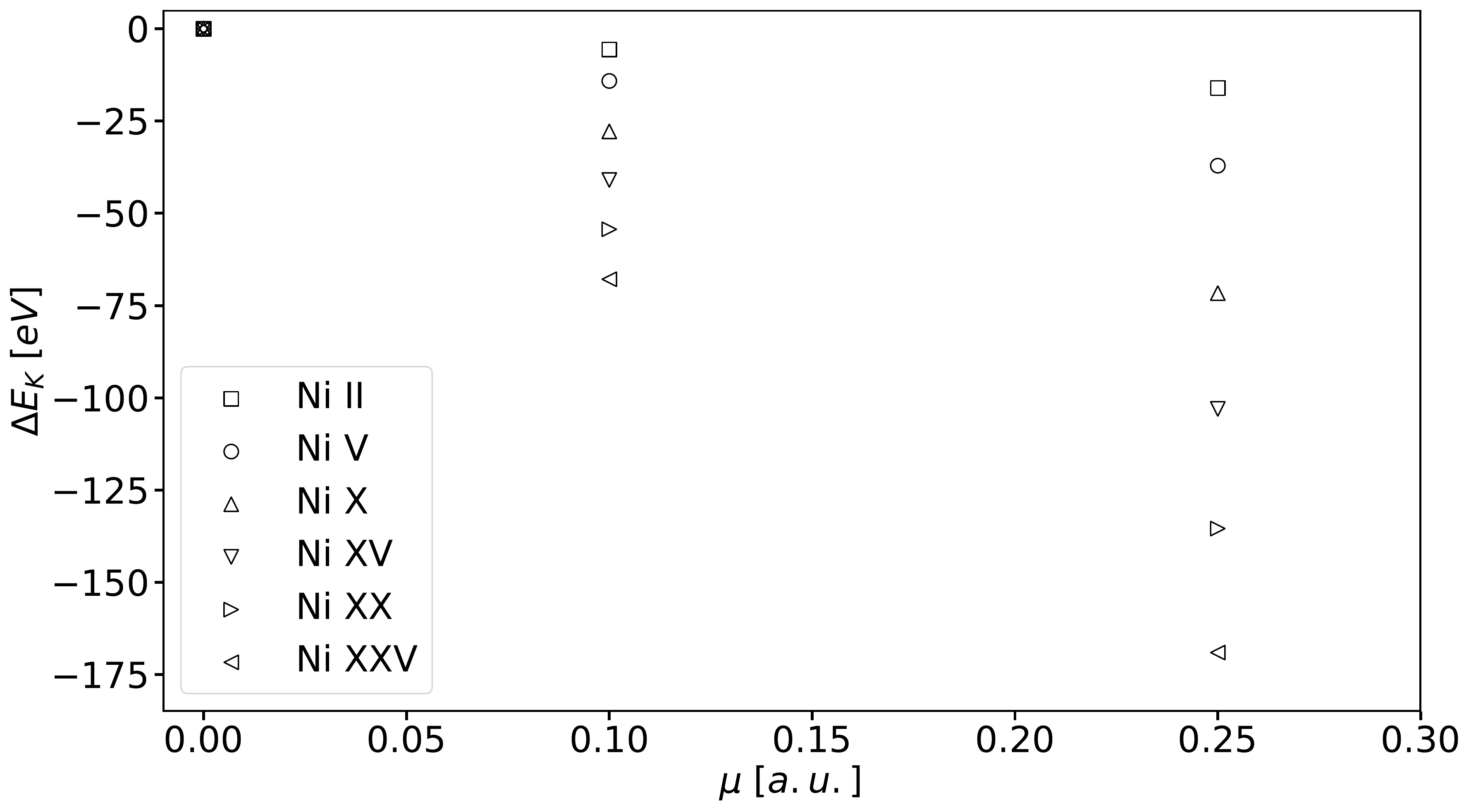}
  \caption{MCDF/RATIP K-threshold shift, $\Delta E_K$ in eV, as a function of the plasma screening parameter, $\mu$ in a.u., in some ions belonging to the nickel isonuclear sequence.} \label{figKTDvsmu_nickel}
\end{figure}

The K-threshold energies, $E_K$ in eV, computed in this work using the MCDF/RATIP method are reported in Tables~\ref{KTC}-\ref{KTNi} for plasma screening parameters $\mu=0$ a.u. (isolated atomic system), $\mu =0.1$ a.u., and $\mu = 0.25$ a.u. for respectively the carbon, silicon, calcium, chromium, and nickel isonuclear sequences. These values are plotted as function of the effective charge in Figure~\ref{figKTvsZeff_nickel} for our example of the nickel ions \ion{Ni}{i}--\ion{Ni}{xxvii}. As for the IPs, all the three curves follow the same trend with systematic lowerings with respect to the isolated atom systems ($\mu = 0$~a.u.). A gradient change is also seen at $Z_{\rm eff} = 19$ corresponding to the closure of the L-shell and opening of the M-shell. In addition, as shown in Figures~\ref{figKTDvsZeff_nickel}--\ref{figKTDvsmu_nickel} for the nickel isonuclear sequence, the K-threshold energy shift, $\Delta E_K = E_K(\mu) - E_K(\mu=0)$ in eV, with respect to the isolated atom condition ($\mu=0$~a.u.) displays a linear dependence on both the effective charge and the plasma screening parameter, in agreement with our results in oxygen (Paper~I) and iron  (Papers~II--IV) ions.

\section{Universal fitting formulae for IP and K-threshold shifts}

The aim of the present work is to refine the universal formulae first provided in our Paper~IV. In order to do so, we have extended the number of different isonuclear sequences used in the fitting of the coefficients multiplying the $\mu~Z_{\rm eff}$ factor by including the MCDF/RATIP IP and K-threshold shifts reported in the previous section for \ion{C}{i}--\ion{C}{v}, \ion{Si}{i}--\ion{Si}{xiii}, \ion{Ca}{i}--\ion{Ca}{xix}, \ion{Cr}{i}--\ion{Cr}{xxiii} and \ion{Ni}{i}--\ion{Ni}{xxvii}. The resulting fitting formulae for the IP and K-threshold lowerings, $\Delta E_0$ and $\Delta E_K ,$ respectively,
in eV, are given below: 
\begin{equation}
\Delta E_0 =~(-26.29\pm0.04)~\mu~Z_{\rm eff} \ ,
\label{uffIP}
\end{equation}
and
\begin{equation}
\Delta E_K =~(-27.27\pm0.05)~\mu~Z_{\rm eff} \ ,
\label{uffKT}
\end{equation}
which are close to the Debye-H\"uckel limit \citep{sp66,crow14}, as shown in Eq.~(\ref{DHlimitIP}) and as was already the case in Paper~IV. The coefficients that appear in Eqs.~(\ref{uffIP}), (\ref{uffKT}) are more constrained by the fits reducing the standard deviation by a factor  of $\sim$2 with respect to those of Paper~IV, where only the iron and oxygen isonuclear sequences were considered in the adjustments. Here, it has to be emphasized that the standard deviations of the fitting parameters are obtained by considering equal weights for the data points used in the fits, \textit{i.e.}, the calculated MCDF/RATIP values are supposed to be exact. The idea is to provide formulae that reproduce our MCDF/RATIP models almost exactly, at least at the resolution of present and future instruments.   

\begin{figure}[!ht]
  \centering
  \includegraphics*[pagebox=mediabox, width=\columnwidth]{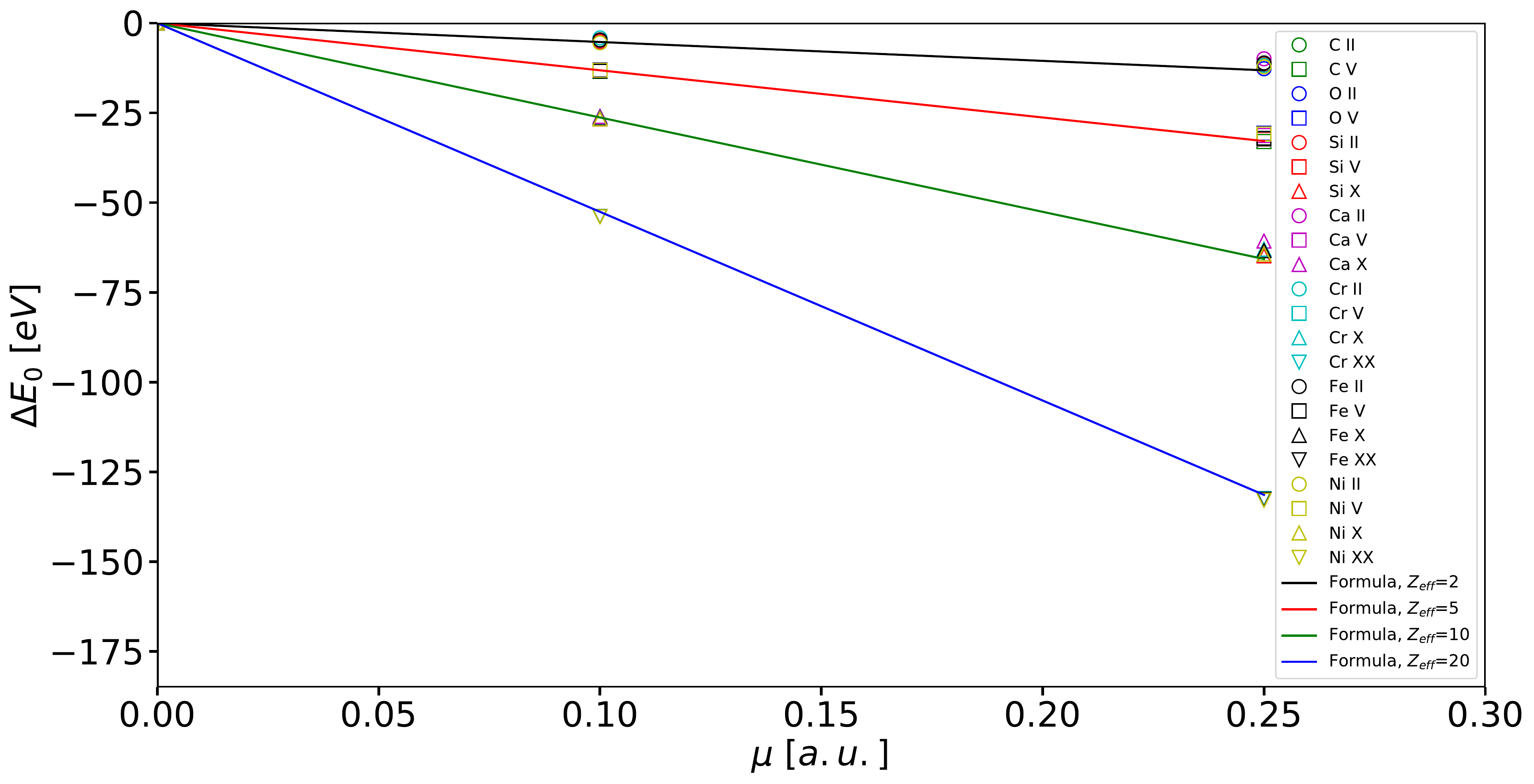}
   \includegraphics*[pagebox=mediabox, width=\columnwidth]{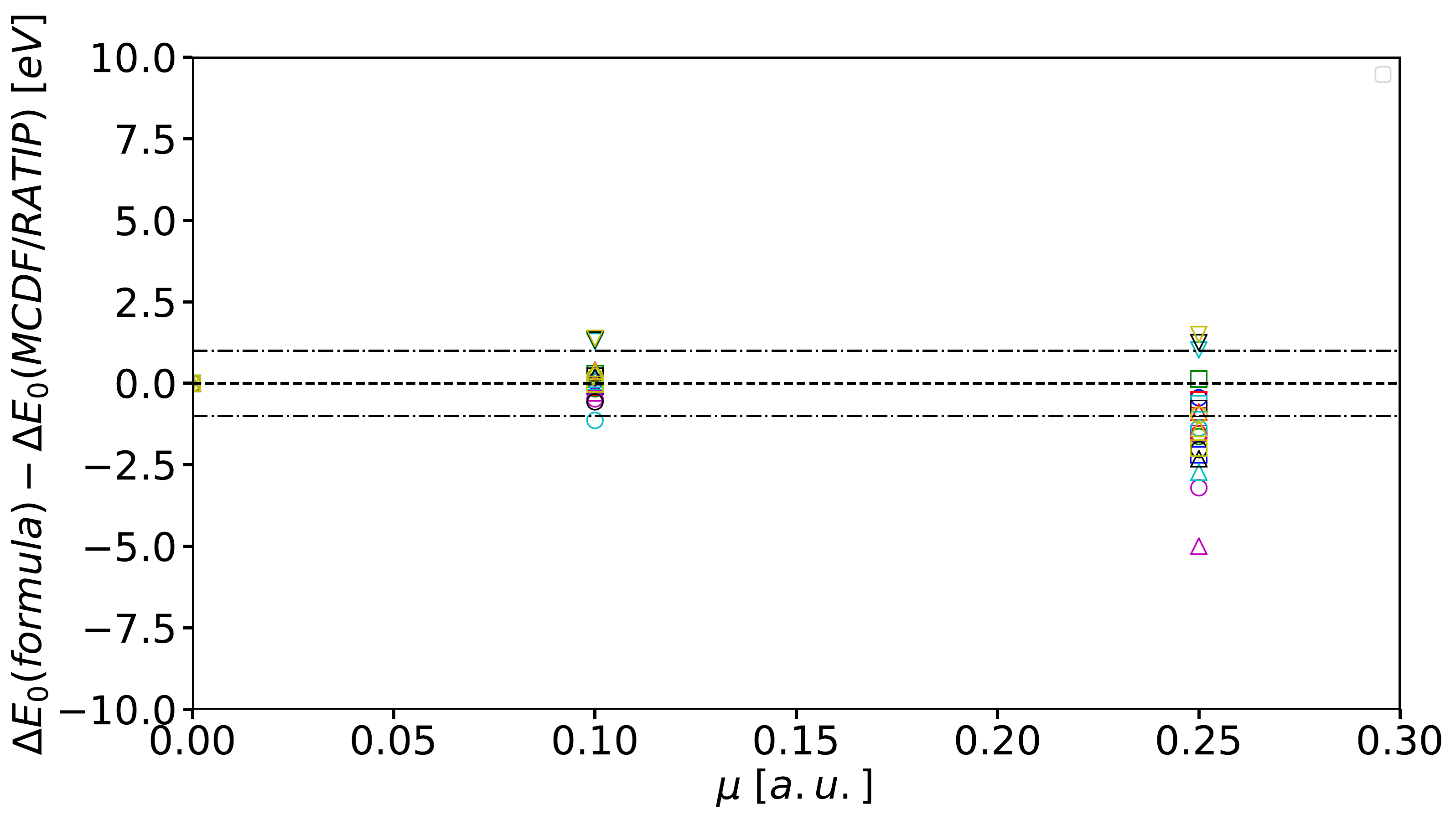}
  \caption{\textit{Top}: Ionization potential lowering, $\Delta E_0$ in eV, as a function of the plasma screening parameter, $\mu$ in a.u., in several abundant ions. Colored lines: Fitting formula (\ref{uffIP}). Colored symbols: MCDF/RATIP method. 
\textit{Bottom}: Differences between universal formula's ionization potential lowerings and MCDF/RATIP values, $(\Delta E_0 (formula) - \Delta E_0 (MCDF/RATIP))$ in eV, as a function of the plasma screening parameter, $\mu$, in several abundant ions. Dashed line: straightline of equality. Dotdashed lines: straight lines of $\pm 1$~eV differences.} \label{figIPDvsmu}
\end{figure}

\begin{figure}[!ht]
  \centering
  \includegraphics*[pagebox=mediabox, width=\columnwidth]{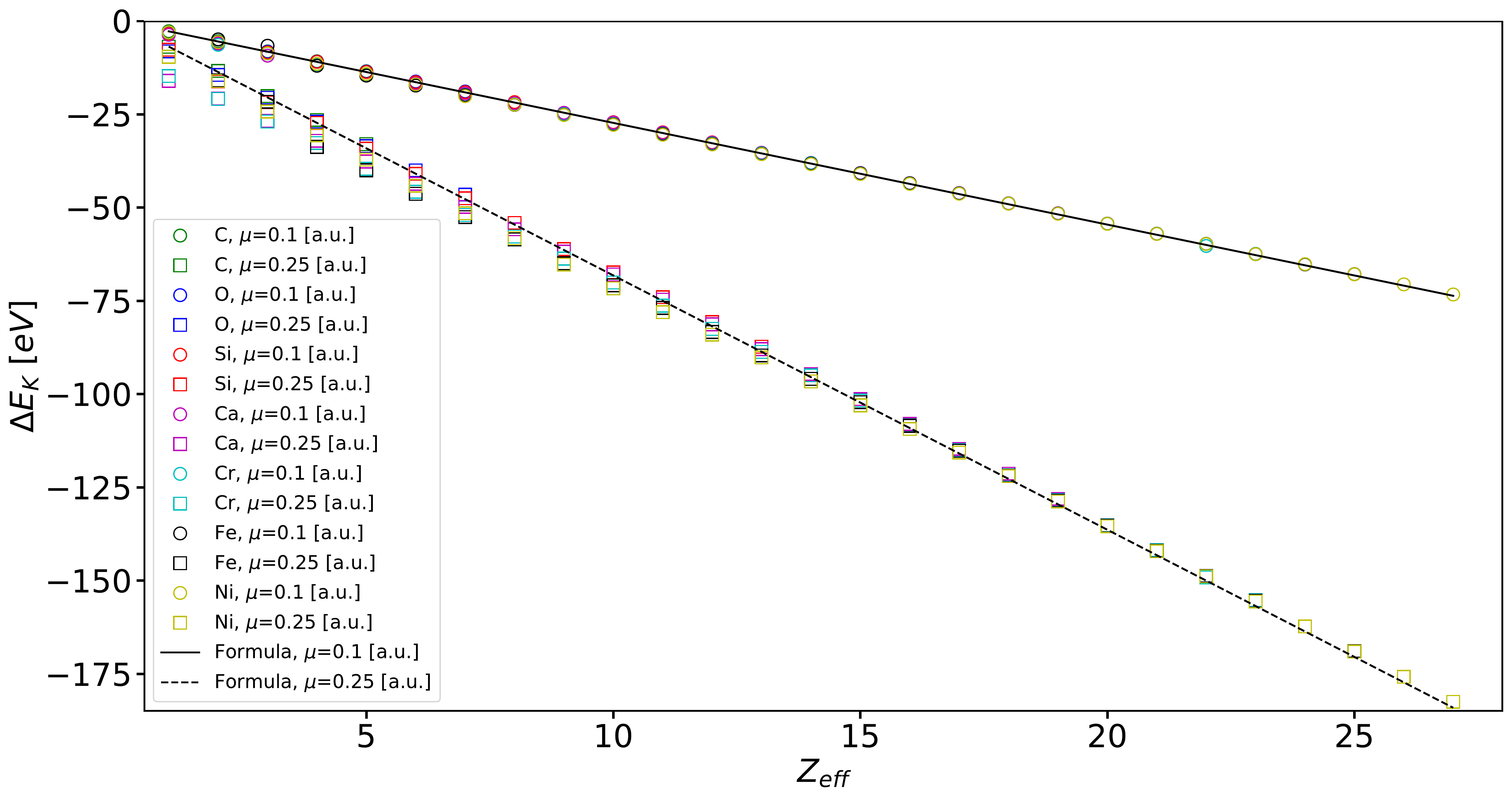}
    \includegraphics*[pagebox=mediabox, width=\columnwidth]{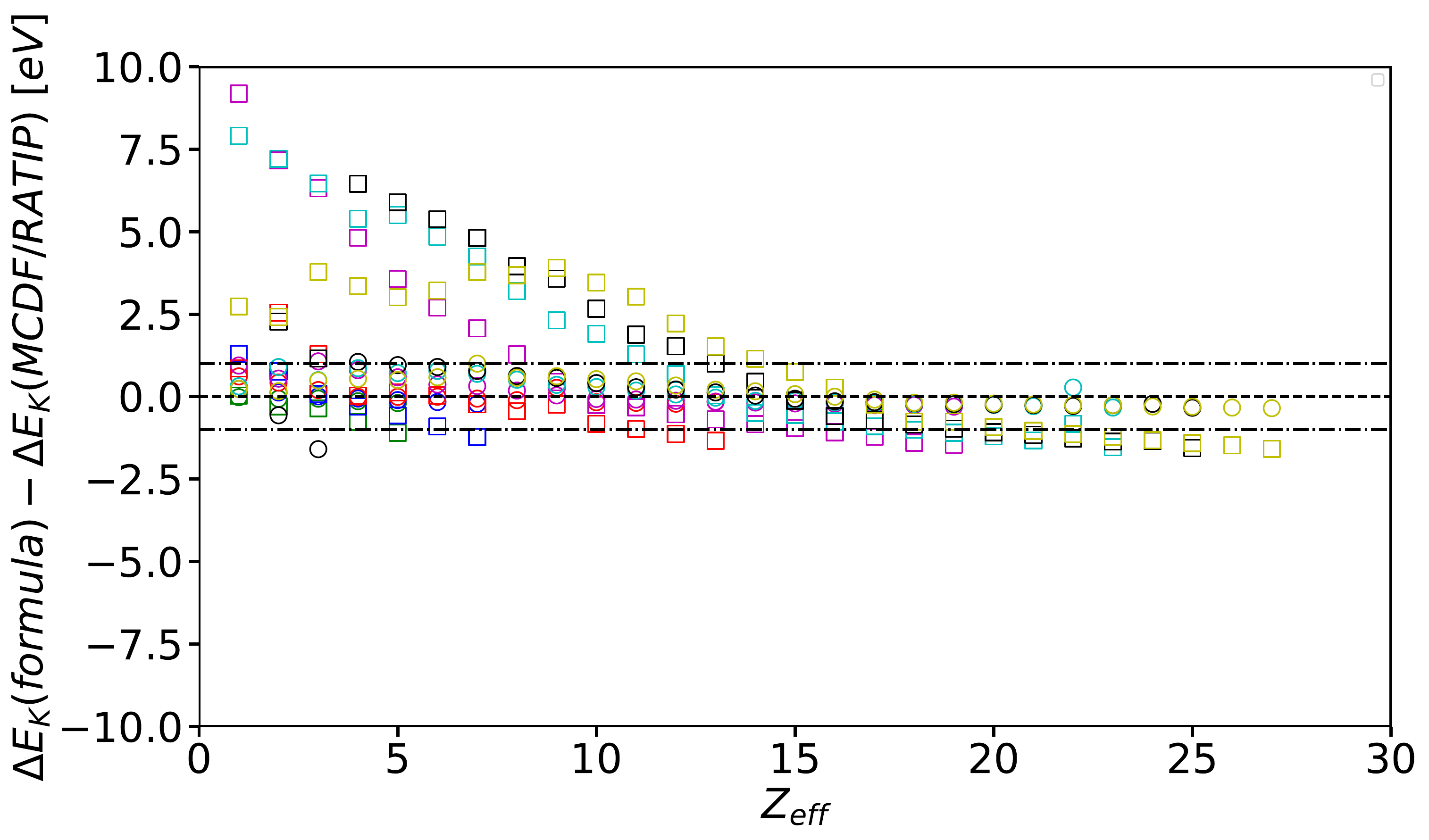}
   \caption{\textit{Top}: K-threshold lowering, $\Delta E_K$ in eV, as a function of the effective charge, $Z_{\rm eff}$, in several abundant ions. Solid line: Fitting formula (\ref{uffKT}) for $\mu = 0.1$~a.u. Dashed line: fitting formula (\ref{uffKT}) for $\mu = 0.25$~a.u. Colored symbols: MCDF/RATIP method. \textit{Bottom}: Differences between universal formula's K-threshold energy lowerings and MCDF/RATIP values, $(\Delta E_K (formula) - \Delta E_K (MCDF/RATIP))$ in eV, as a function of the effective charge, $Z_{\rm eff}$, in several abundant ions. Dashed line: straight line of equality. Dotted-dashed lines: Straight lines of $\pm 1$~eV differences.} \label{figKTDvsZeff}
\end{figure}

\begin{figure}[!ht]
  \centering
  \includegraphics*[pagebox=mediabox, width=\columnwidth]{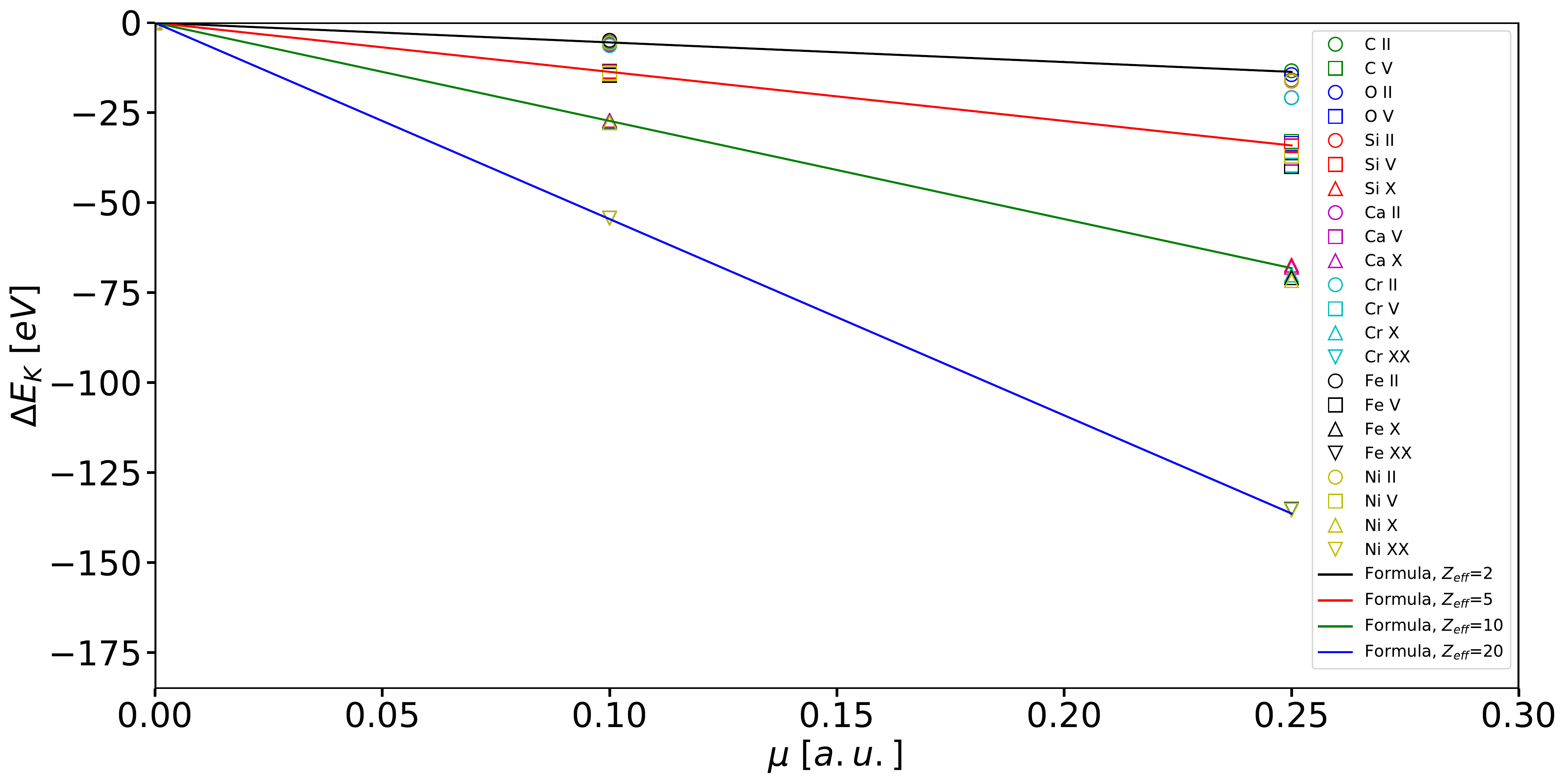}
    \includegraphics*[pagebox=mediabox, width=\columnwidth]{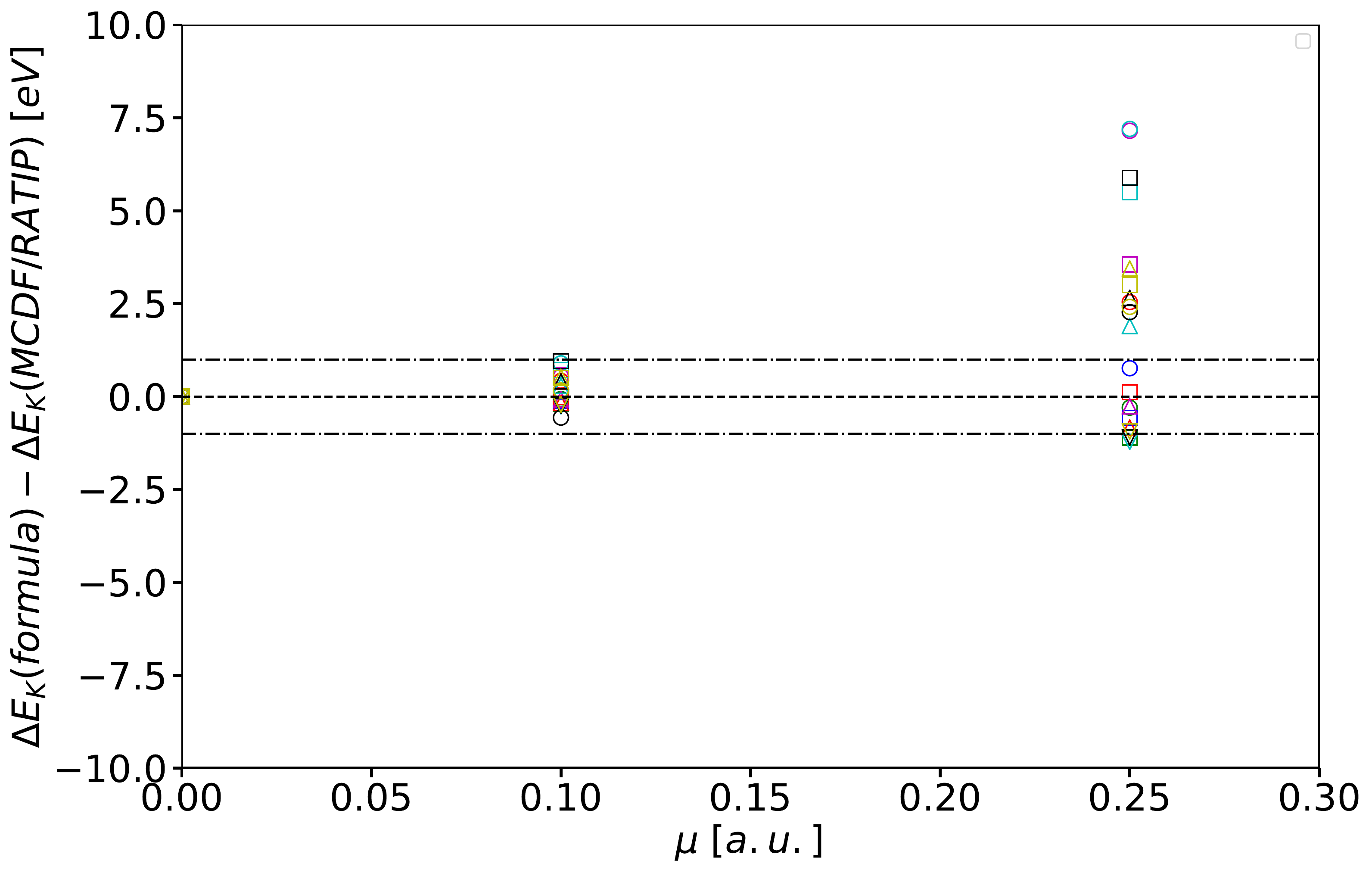}
  \caption{\textit{Top}: K-threshold lowering, $\Delta E_K$ in eV, as a function of the plasma screening parameter, $\mu$ in a.u., in several abundant ions. Colored lines: fitting formula (\ref{uffKT}). Colored symbols: MCDF/RATIP method.
  \textit{Bottom}: Differences between universal formula's K-threshold energy lowerings and MCDF/RATIP values, $(\Delta E_K (formula) - \Delta E_K (MCDF/RATIP))$ in eV, as a function of the plasma screening parameter, $\mu$, in several abundant ions. Dashed line: Straight line of equality. Dotted-dashed lines: Straight lines of $\pm 1$~eV differences.} \label{figKTDvsmu}
\end{figure}

\begin{figure}[!ht]
  \centering
  \includegraphics*[pagebox=mediabox, width=\columnwidth]{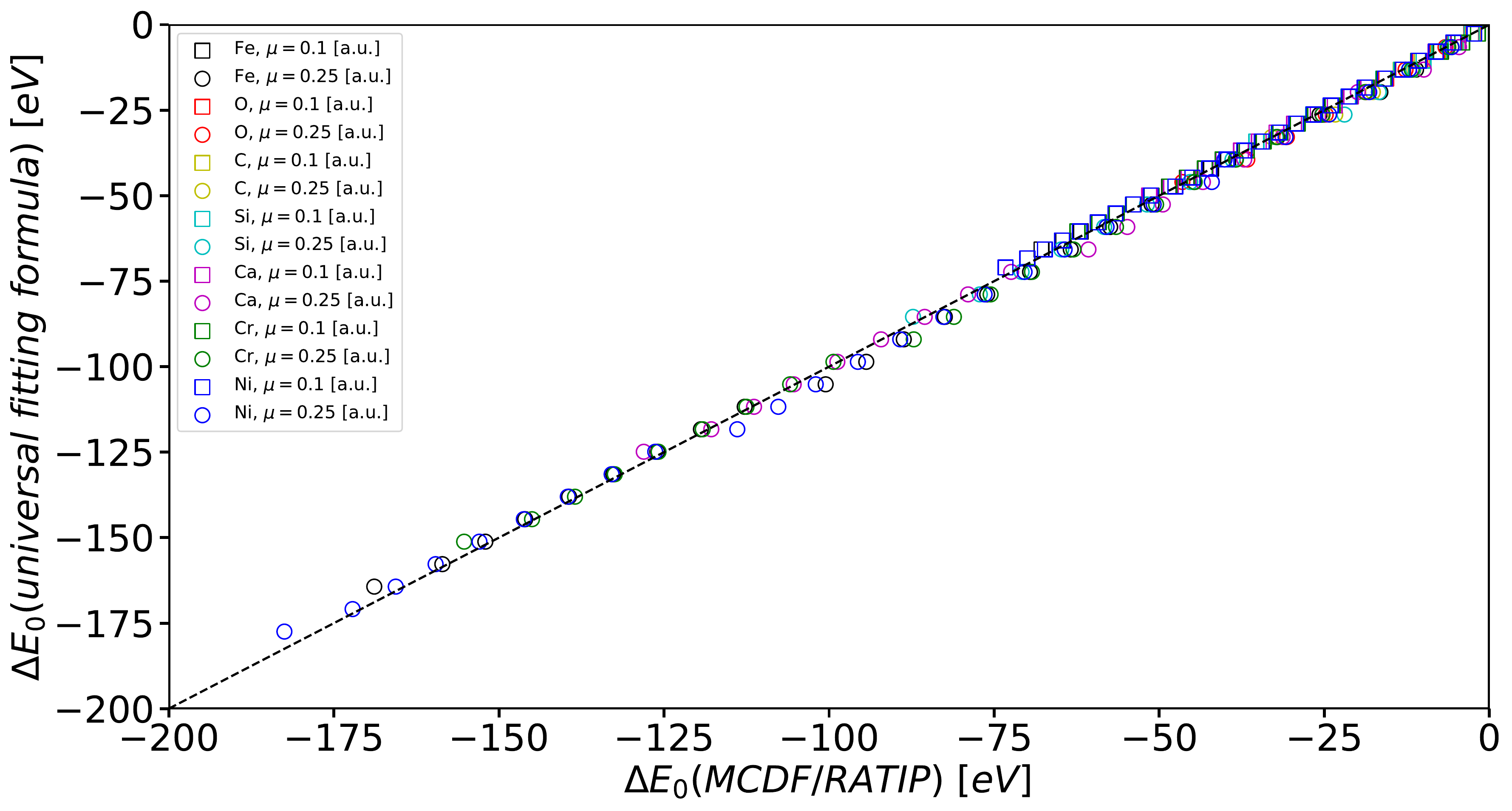}
    \includegraphics*[pagebox=mediabox, width=\columnwidth]{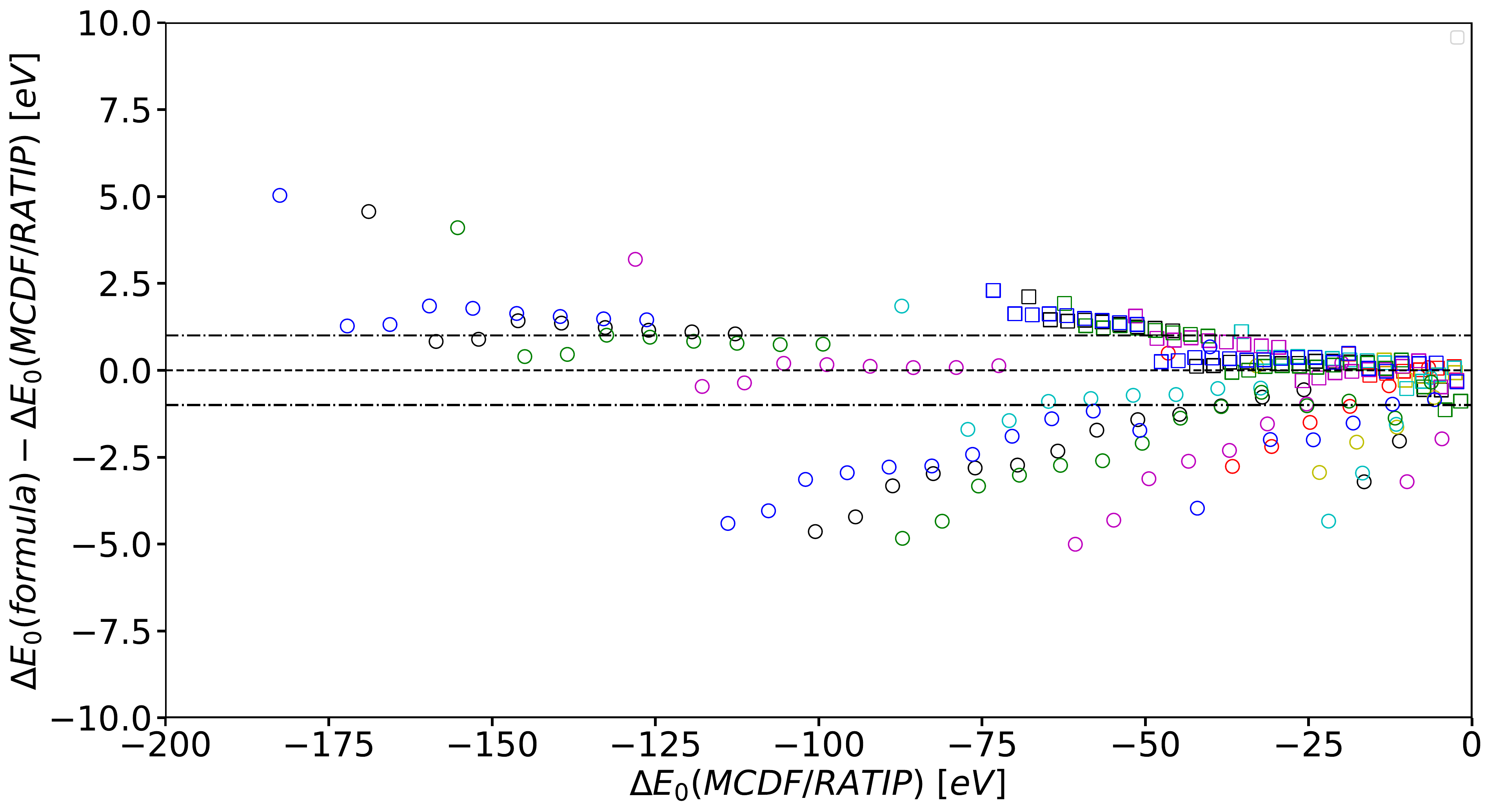}
  \caption{\textit{Top}: Comparison between the ionization potential shifts obtained by the fitting formula (\ref{uffIP}) and the MCDF/RATIP values computed in this work. Dashed line: Straight line of equality.
  \textit{Bottom}: Differences between universal formula's ionization potential lowerings and MCDF/RATIP values, $(\Delta E_0 (formula) - \Delta E_0 (MCDF/RATIP))$ in eV, as a function of $\Delta E_0 (MCDF/RATIP)$ in several abundant ions. Dashed line: Straight line of equality. Dotted-dashed lines: Straight lines of $\pm 1$~eV differences.} \label{figIPDcomp}
\end{figure}

\begin{figure}[!ht]
  \centering
  \includegraphics*[pagebox=mediabox, width=\columnwidth]{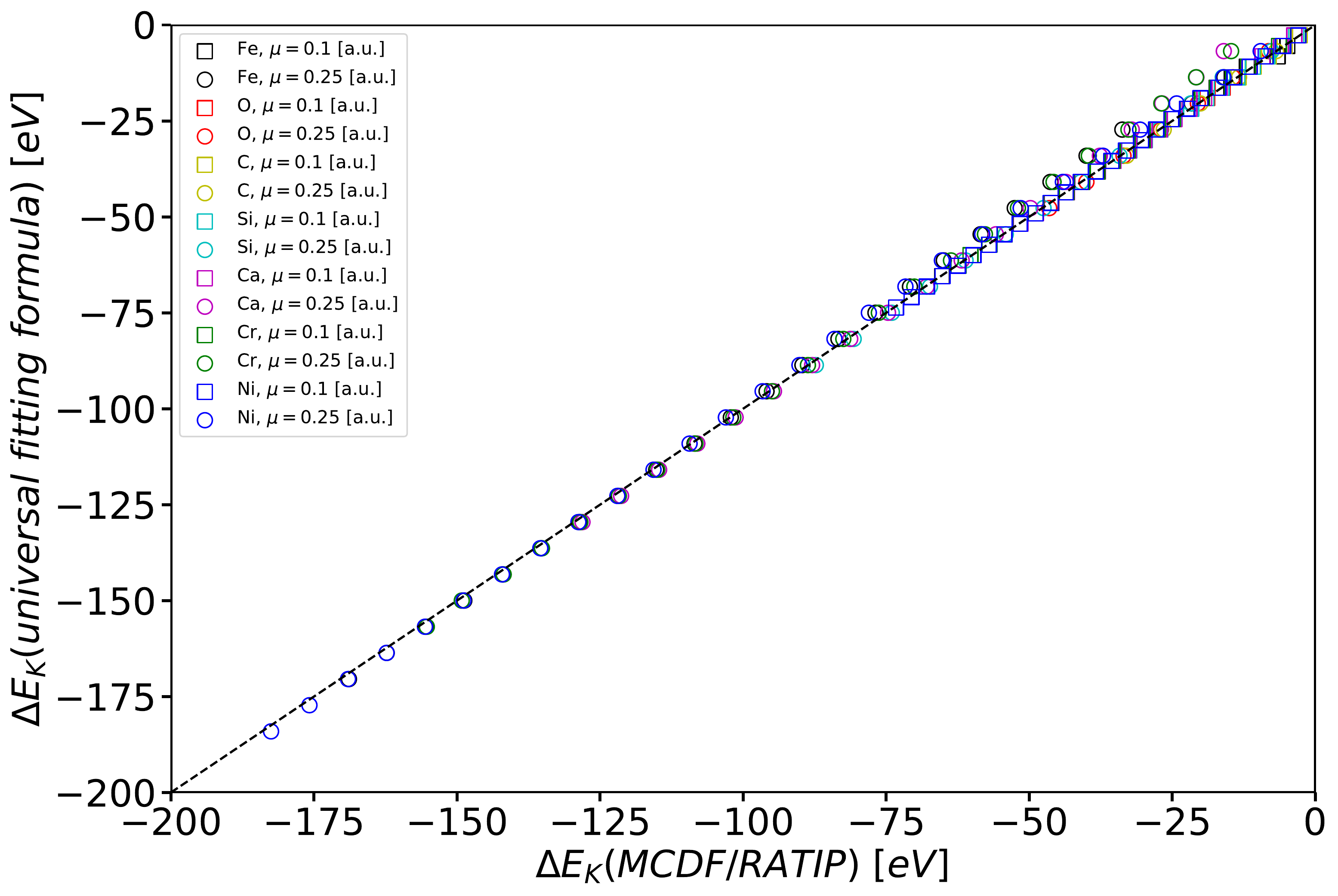}
   \includegraphics*[pagebox=mediabox, width=\columnwidth]{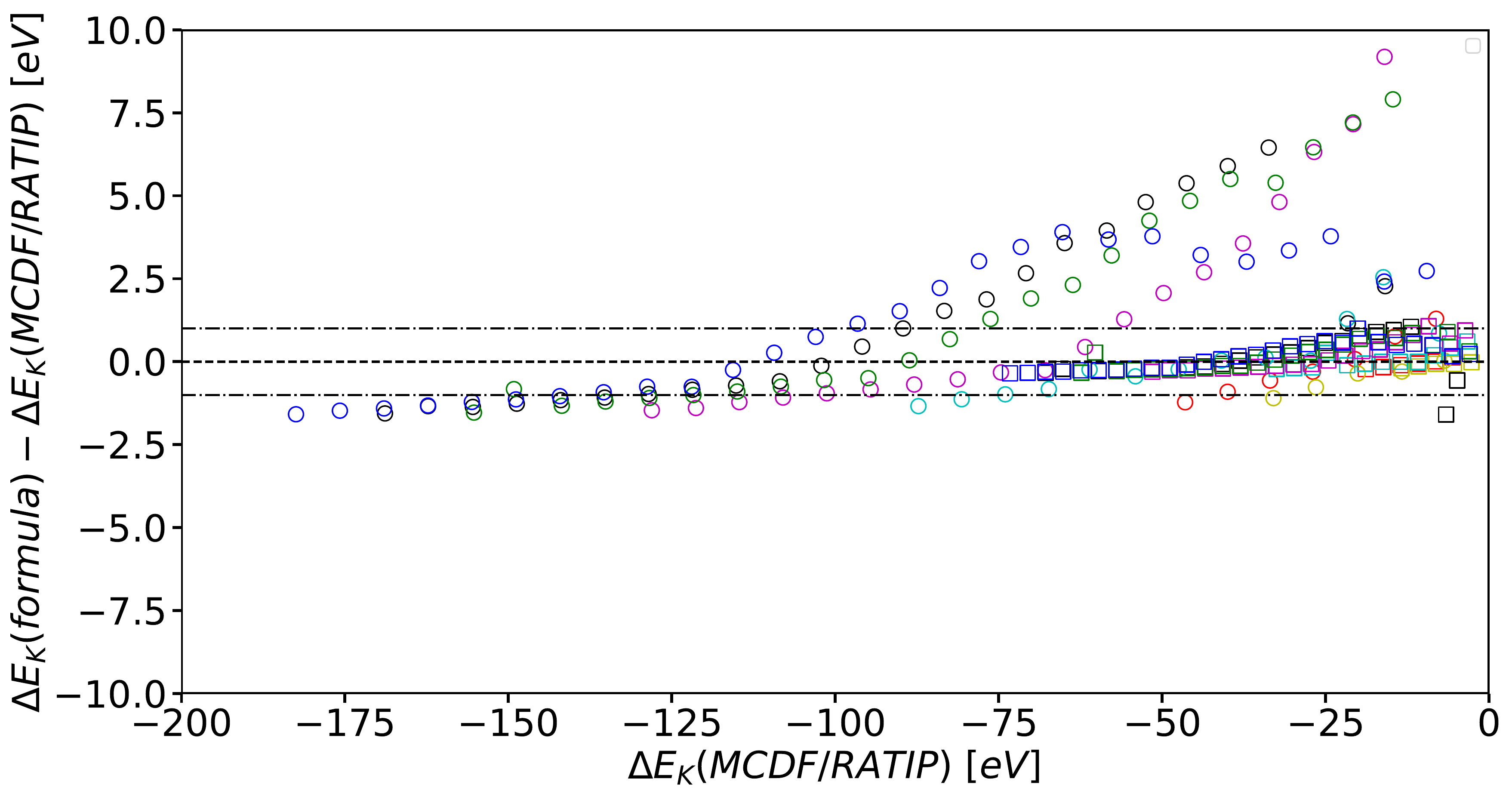}
  \caption{\textit{Top}: Comparison between the K-threshold energy shifts obtained by the fitting formula (\ref{uffKT}) and the MCDF/RATIP values computed in this work. Dashed line: Straight line of equality.
  \textit{Bottom}: Differences between universal formula's K-threshold energy lowerings and MCDF/RATIP values, $(\Delta E_K (formula) - \Delta E_K (MCDF/RATIP))$ in eV, as a function of $\Delta E_K (MCDF/RATIP)$ in several abundant ions. Dashed line: Straight line of equality. Dotted-dashed lines: Straight lines of $\pm 1$~eV differences.} \label{figKTDcomp}
\end{figure}

In Figures~\ref{figIPDvsZeff} and~\ref{figIPDvsmu}, we display the resulting fits to the MCDF/RATIP IP lowerings computed in this work and in Papers~I--IV, respectively, as a function of the effective charge and of the plasma screening parameter. In Figure~\ref{figIPDvsmu}, we restrict the plot to a sample of effective charges for the sake of clarity.
Figures~\ref{figKTDvsZeff} and \ref{figKTDvsmu} are the equivalents for the K-threshold energy shifts. In Figure~\ref{figIPDcomp} and Figure~\ref{figKTDcomp}, comparisons between computed and fitted values are presented for $\Delta E_0$ and $\Delta E_K,$ respectively. The latter all show a sound reliability, although some minor scatter appears especially for $\mu = 0.25$~a.u. In order to investigate further, the differences between the predicted values given by the formulae (\ref{uffIP}) and (\ref{uffKT}) and the MCDF/RATIP values are plotted versus the effective charge, the plasma screening parameter and the MCDF/RATIP value in the bottom panels of, respectively, Figures~\ref{figIPDvsZeff} and \ref{figKTDvsZeff} as well as Figures~\ref{figIPDvsmu} and \ref{figKTDvsmu}  as well as Figures~\ref{figIPDcomp} and \ref{figKTDcomp}, for the IP and K-threshold lowerings, respectively. As we see, the differences range from up to 5 eV to less than 1 eV in the bottom panel of Figure~\ref{figIPDvsZeff} and from up to 9 eV to less than 1 eV in the bottom panel of Figure~\ref{figKTDvsZeff} with bigger differences for $\mu = 0.25$~a.u. and with agreements within $\sim$2 eV for $\mu = 0.1$~a.u. The average of the absolute differences are 1.2$\pm$1.2~eV and 1.1$\pm$1.6~eV respectively for the IP and K-threshold lowerings. Moreover, these average differences drop to  0.6$\pm$0.5~eV and 0.3$\pm$0.3~eV, respectively, when only the values for $\mu=0.1$~a.u are considered and increase up to 1.8$\pm$1.3~eV and 2.0$\pm$1.9~eV, respectively, when only the values for $\mu = 0.25$~eV are retained this time. All the figures are to be compared to the resolve energy scale of the microcalorimeter onboard the future \textit{XRISM} mission, that is, 2~eV with a goal of 1~eV \citep{mil20}. In that respect, the IP and K-threshold lowering differences for $\mu=0.1$~a.u shown in the bottom panels of Figures~\ref{figIPDvsZeff}-\ref{figKTDcomp} are all within $\pm \sim 2$~eV, which is comparable to the \textit{XRISM} resolve energy scale. In contrast, a large number of the values for $\mu=0.25$~a.u are greater than this scale, as notably seen in the bottom panels of Figures~\ref{figIPDvsmu} and \ref{figKTDvsmu}.

In trying to reproduce  the direct MCDF/RATIP calculations more accurately with the fitted formulae, we used a different fitted formula for each shell filling, namely:\ K-shell ions (\ion{C}{v}, \ion{O}{vi}, \ion{Si}{xiii}, \ion{Ca}{xix}, \ion{Cr}{xxiii}, \ion{Fe}{xxv,} and \ion{Ni}{xxvii}), L-shell ions (\ion{C}{i}--\ion{C}{iv}, \ion{O}{i}--\ion{O}{v}, \ion{Si}{v}--\ion{Si}{xii}, \ion{Ca}{xi}--\ion{Ca}{xviii}, \ion{Cr}{xv}--\ion{Cr}{xxii}, \ion{Fe}{xvii}--\ion{Fe}{xxiv,} and \ion{Ni}{xix}--\ion{Ni}{xxvi}), and M-shell ions (\ion{Si}{i}--\ion{Si}{iv}, \ion{Ca}{iii}--\ion{Ca}{x}, \ion{Cr}{i}--\ion{Cr}{xiv}, \ion{Fe}{ii}--\ion{Fe}{xvi,} and \ion{Ni}{i}--\ion{Ni}{xviii}). The resulting fitted formulae are the following:

\begin{figure}[!ht]
  \centering
  \includegraphics*[pagebox=mediabox, width=\columnwidth]{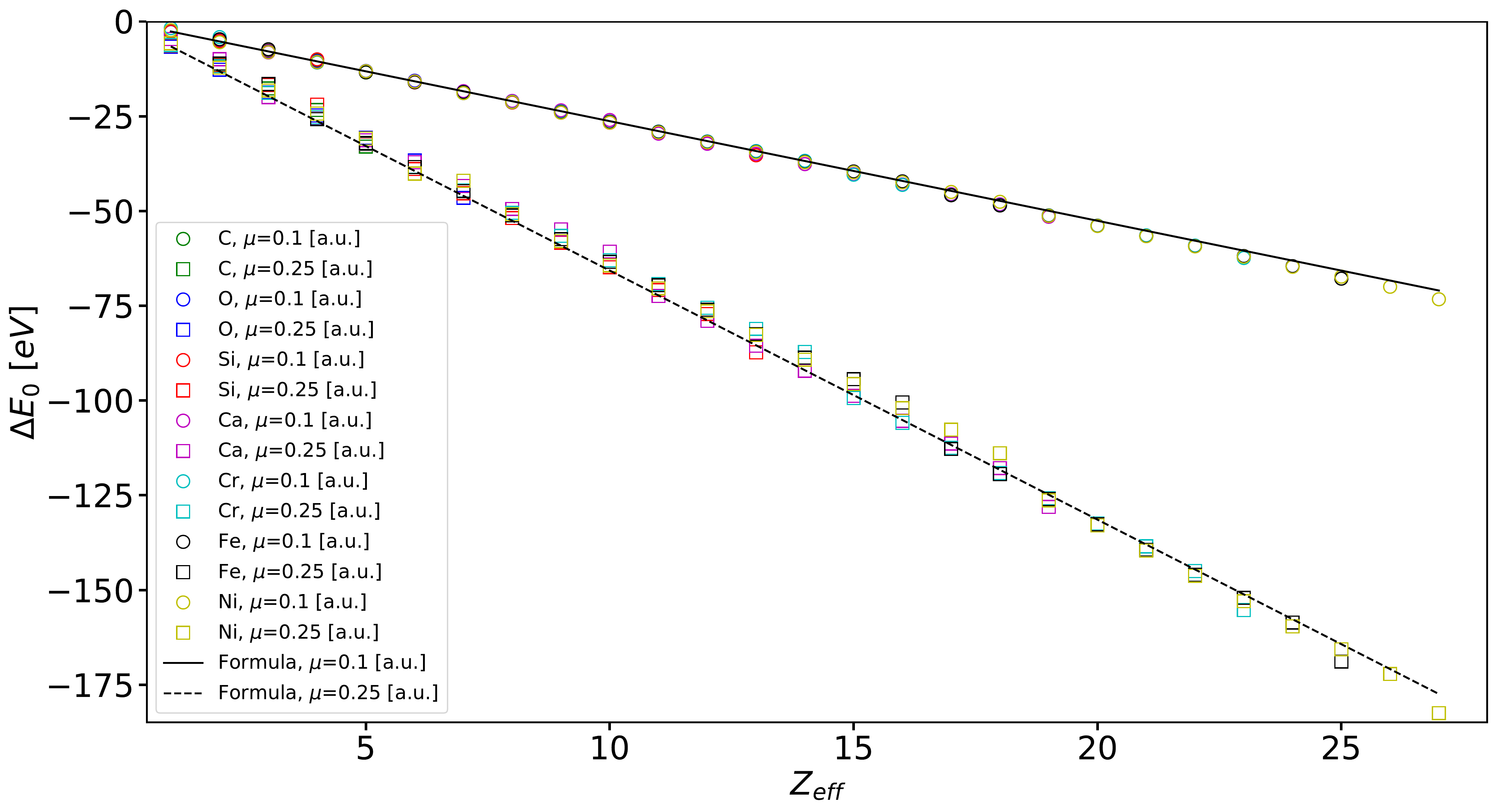}
  \includegraphics*[pagebox=mediabox, width=\columnwidth]{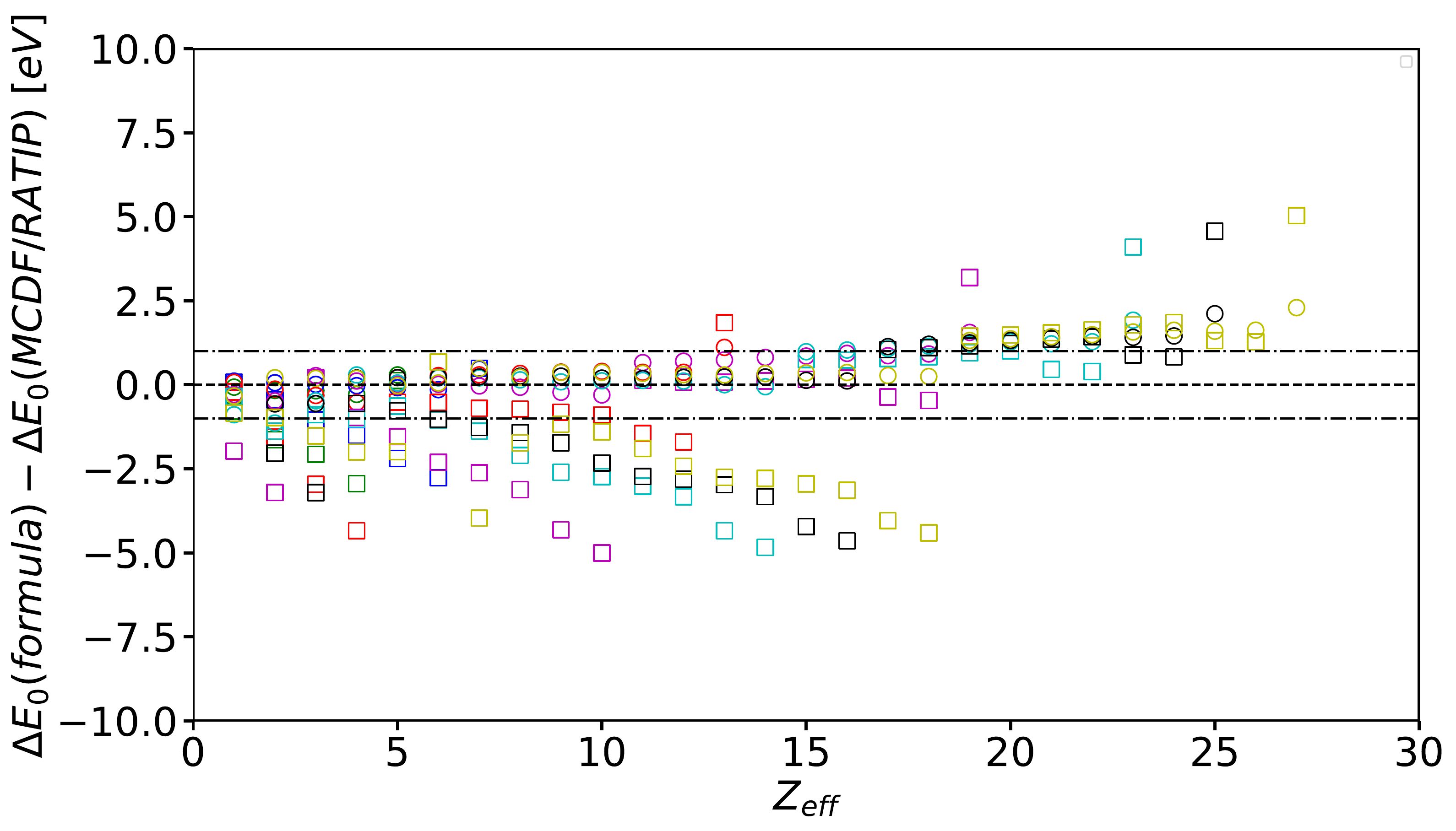}
  \caption{\textit{Top}: Ionization potential lowering, $\Delta E_0$ in eV, as a function of the effective charge, $Z_{\rm eff}$, in several abundant ions. Solid line: fitting formula (\ref{uffIP}) for $\mu = 0.1$~a.u. Dashed line: fitting formula (\ref{uffIP}) for $\mu = 0.25$~a.u. Colored symbols: MCDF/RATIP method. 
\textit{Bottom}: Differences between universal formula's ionization potential lowerings and MCDF/RATIP values, $(\Delta E_0 (formula) - \Delta E_0 (MCDF/RATIP))$ in eV, as a function of the effective charge, $Z_{\rm eff}$, in several abundant ions. Dashed line: straightline of equality. Dotdashed lines: straight lines of $\pm 1$~eV differences. } \label{figIPDvsZeff}
\end{figure}

\begin{equation}
\Delta E_0^{(K)} =~(-27.01\pm0.03)~\mu~Z_{\rm eff} \ ,
\label{uffIPK}
\end{equation}
and
\begin{equation}
\Delta E_K^{(K)} =~(-27.01\pm0.03)~\mu~Z_{\rm eff} \ ,
\label{uffKTK}
\end{equation}
for the K-shell ions;
\begin{equation}
\Delta E_0^{(L)} =~(-26.50\pm0.03)~\mu~Z_{\rm eff} \ ,
\label{uffIPL}
\end{equation}
and
\begin{equation}
\Delta E_K^{(L)} =~(-27.07\pm0.01)~\mu~Z_{\rm eff} \ ,
\label{uffKTL}
\end{equation}
for the L-shell ions;
\begin{equation}
\Delta E_0^{(M)} =~(-25.41\pm0.06)~\mu~Z_{\rm eff} \ ,
\label{uffIPM}
\end{equation}
and
\begin{equation}
\Delta E_K^{(M)} =~(-27.92\pm0.11)~\mu~Z_{\rm eff} \ ,
\label{uffKTM}
\end{equation}
for the M-shell ions. The corresponding differences with respect to the MCDF/RATIP values are all within 1~eV for formulae~(\ref{uffIPK}) and (\ref{uffKTK}) for the K-shell ions. For the L-shell ions, they are also all within $\sim$ 1 eV for K-threshold lowerings given by formula~(\ref{uffKTL}) but they scatter more, namely, within $\sim$3~eV for the IP lowerings (Eq.~(\ref{uffIPL}), essentiallydue to  the $\mu = 0.25$~a.u. values in \ion{C}{ii}--\ion{C}{iv}, \ion{O}{iv}--\ion{O}{vi,} and \ion{Si}{xi}--\ion{Si}{xii}.  Regarding the M-shell ions, formulae~(\ref{uffIPM}) and (\ref{uffKTM})
predict values that differ from the MCDF/RATIP calculations by up to $\sim$7.5 eV (in \ion{Cr}{i} under conditions parametrized by $\mu = 0.25$~a.u.) for the K-threshold lowerings and up to $\sim$3~eV (in \ion{Si}{iv} under conditions parametrized by $\mu = 0.25$~a.u.) for the IP lowerings. These do not therefore solve the problem of better accuracy compared to the resolve energy scale attained by future x-ray space telescopes. In conclusion, the universal formulae~(\ref{uffIP}) and (\ref{uffKT}) are of equivalent accuracy then the shell specific formulae~(\ref{uffIPK}--\ref{uffKTM}) but of more of practical use for the astrophysical plasma modeling codes such as {\sc xstar} \citep{bau01,kal01,kal21,men21}. 

\begin{figure}[!ht]
  \centering
  \includegraphics*[pagebox=mediabox, width=\columnwidth]{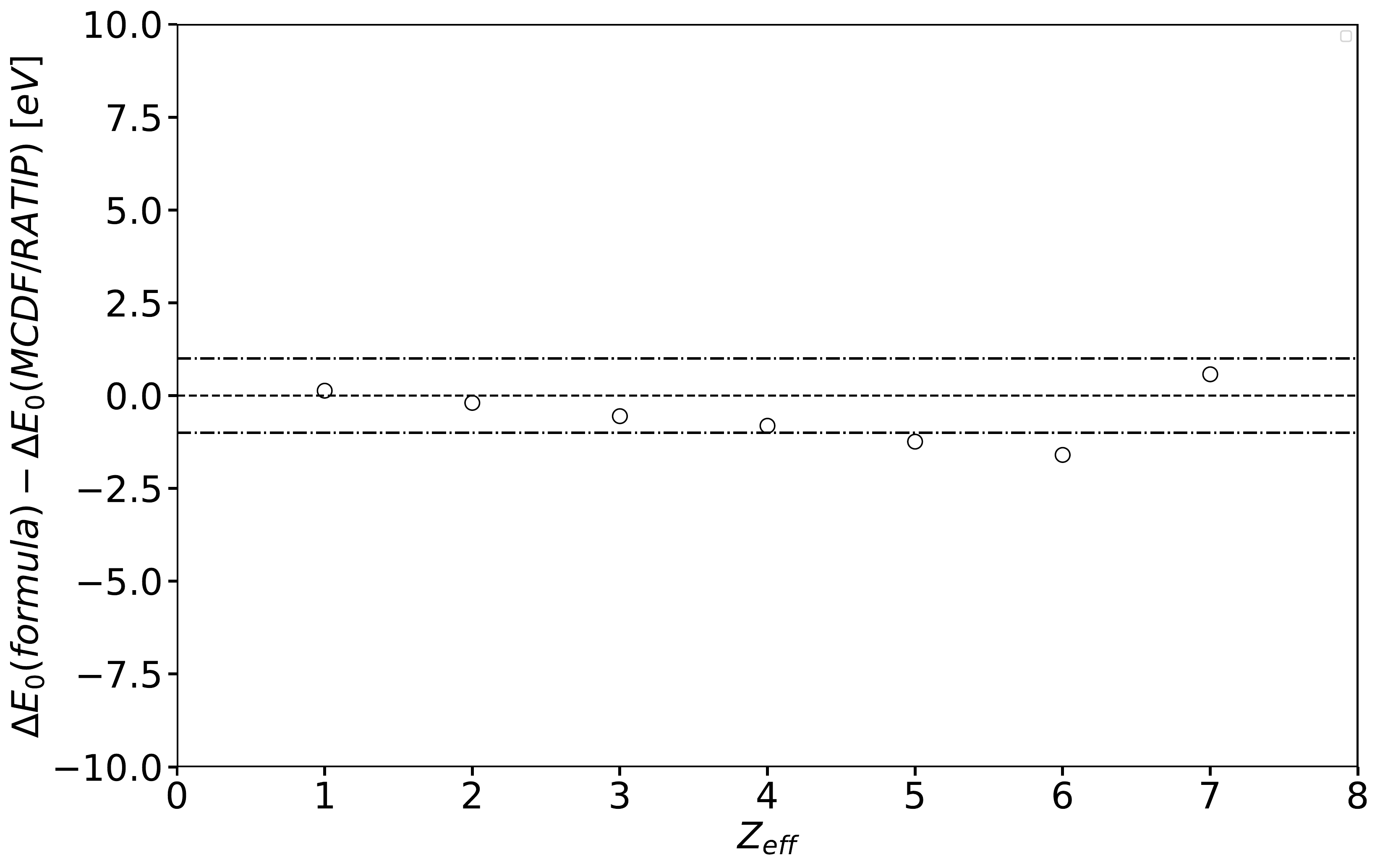}
  \includegraphics*[pagebox=mediabox, width=\columnwidth]{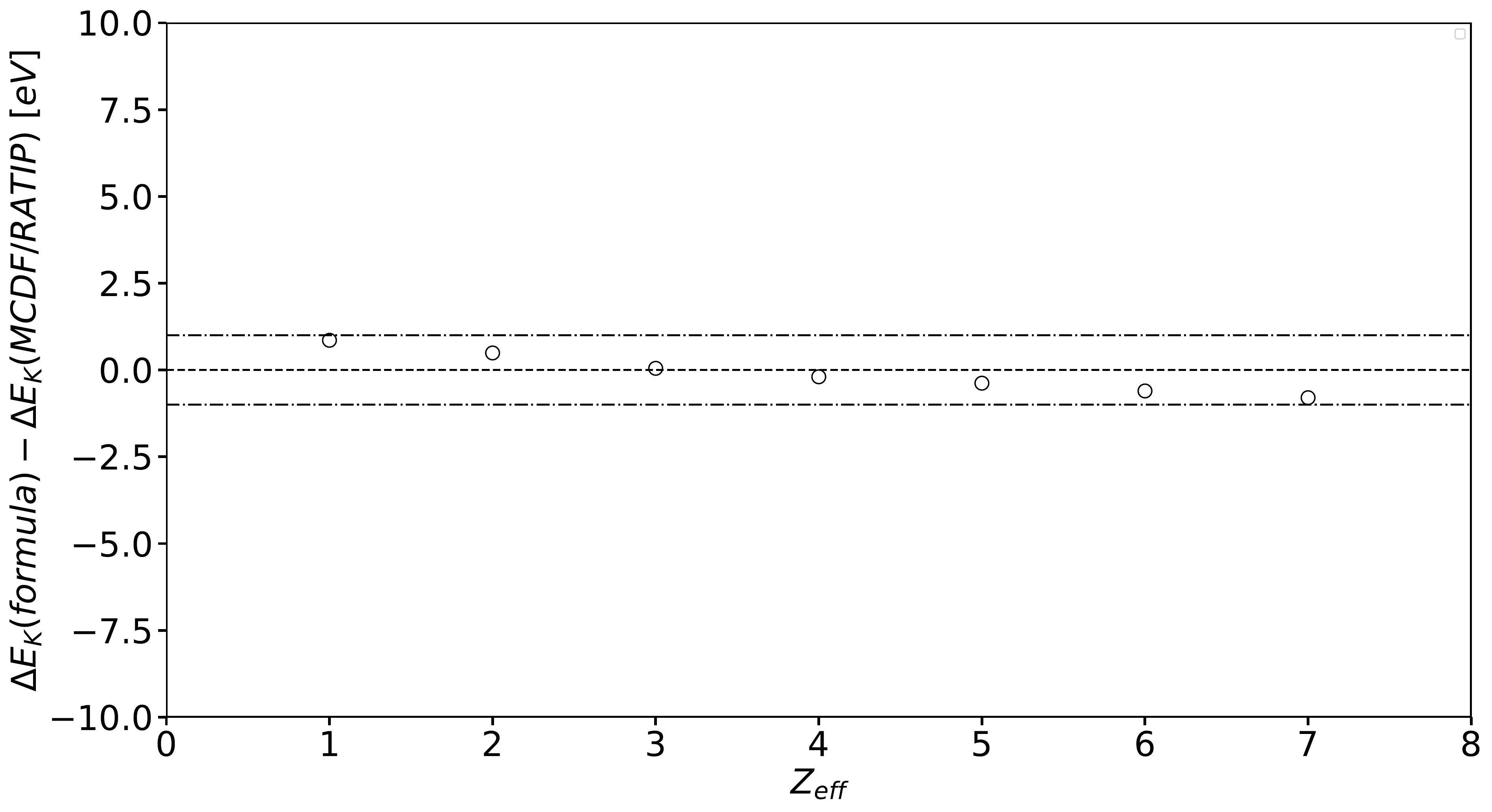}
  \caption{\textit{Top}: Differences between universal formula's ionization potential lowerings and MCDF/RATIP values, $(\Delta E_0 (formula) - \Delta E_0 (MCDF/RATIP))$ in eV, as a function of the effective charge, $Z_{\rm eff}$, in the oxygen isonuclear sequence for $\mu = 0.2$~a.u. Dashed line: Straight line of equality. Dotted-dashed lines: Straight lines of $\pm 1$~eV differences.
  \textit{Bottom}: Differences between universal formula's K-threshold energy lowerings and MCDF/RATIP values, $(\Delta E_K (formula) - \Delta E_K (MCDF/RATIP))$ in eV, as a function of the effective charge, $Z_{\rm eff}$, the oxygen isonuclear sequence for $\mu = 0.2$~a.u. Dashed line: Straight line of equality. Dotted-dashed lines: Straight lines of $\pm 1$~eV differences.} \label{figIPKTDcomp_O}
\end{figure}

Finally, we used the MCDF/RATIP values calculated in the oxygen isonuclear sequence for $\mu = 0.2$~a.u. in Paper~IV for comparison with the predictions given by the fitting formulae, namely,  Eqs. (\ref{uffIP}), (\ref{uffKT}), and we find a good agreement, as can be seen in Figure~\ref{figIPKTDcomp_O}. Here, the absolute differences range from $\sim$0.1~eV (in \ion{O}{i}) to $\sim$1.6~eV (in \ion{O}{vi}) for the IP lowerings and from $\sim$0.05~eV (in \ion{O}{iii}) to $\sim$0.9~eV (in \ion{O}{i}) for the K-threshold lowerings. At this point, we can conclude that our fitted formulae reproduce our MCDHF/RATIP models with an accuracy comparable to the \textit{XRISM} resolve energy scale for plasma screening parameter values $\mu \leq 0.2$~a.u. 

\section{Summary and conclusions}

We first proposed universal formulae applicable to all relevant elements in order to predict IP and K-threshold lowerings that are due to the screening effect on atomic systems embedded in plasmas (our Paper~IV). In an attempt to improve them, MCDF/RATIP calculations of IP and K-threshold energies have been carried out in the carbon, silicon, calcium, chromium, and nickel isonuclear sequences in order to increase the number of data points that can be used in the fits. The theoretical models used are based on the multiconfiguration Dirac-Fock method that includes a Debye-H\"uckel model potential to take into account the effects of the plasma screening on atomic structures. In each sequence, the MCDF/RATIP IP and K-threshold downshifts show similar linear trends on both the ion effective charge and the plasma screening parameter as the ones previously reported in oxygen and iron isonuclear sequences in, respectively, our Paper~I and our Papers~II--IV. 

Concerning the improved universal formulae, namely those of Eqs.~(\ref{uffIP}) and (\ref{uffKT}), the coefficients of proportionality closed to the Debye-H\"uckel limit are more constrained by the fits, reducing the standard deviations by a factor two with respect to those determined in Paper~IV. Nonetheless, differences with the MCDF/RATIP values are still found to be sizable compared to the \textit{XRISM} resolve energy scale (2~eV) in certain ions near the neutral end of isonuclear sequences or in atomic structures close to closed-shell atoms and for extreme plasma conditions, that is, for $\mu = 0.25$~a.u. Actually, discontinuities appear in the different trends and increase with $\mu$, reflecting the shell structure of the different atomic systems. Considering these discontinuities by using different fitting curves for each shell filling does not solve this issue of accuracy. 

In conclusion, as explicit calculations of the atomic structures for each ion of each element under different plasma conditions is impractical, we still recommend the use of universal formulae, namely, Equations (\ref{uffIP}--\ref{uffKT}) for predicting the IP and K-threshold lowerings in plasma modeling codes such as {\sc xstar} \citep{bau01,kal01,kal21,men21}, although their comparatively moderate to low accuracies in certain cases under extreme plasma conditions characterized by $\mu > 0.2$~a.u. (especially for the K-thresholds) may affect the predicted opacities.

\begin{acknowledgements}
JD is Research Fellow of the Belgian American Educational Foundation Inc. (BAEF) while PP and PQ are, respectively, Research Associate and Research Director of the Belgian Fund for Scientific Research (F.R.S.-FNRS). Financial supports from these organizations, as well as from the NASA Astrophysics Research and Analysis Program (grant 80NSSC17K0345) are gratefully acknowledged. JAG acknowledges support from the Alexander von Humboldt Foundation.
\end{acknowledgements}

\begin{appendix}

\section{Computed ionization potentials for different plasma screening parameter values and comparison with the NIST recommended values}

\begin{table}[!ht]
  \caption{Computed ionization potentials for \ion{C}{i}--\ion{C}{v} as a function of the plasma screening parameter $\mu$ (a.u.). Spectroscopic values (NIST) are also listed for comparison. \label{IPC}}
  \centering
        \begin{tabular}{lrrrr}
                \hline \hline
                \noalign{\smallskip}
                Ion    & \multicolumn{4}{c}{IP (eV)} \\
                \cline{2-5}
                \noalign{\smallskip}
                &  NIST$^a$ & $\mu = 0.0$ & $\mu = 0.1$ & $\mu = 0.25$ \\
                \hline
                \noalign{\smallskip}
                \ion{C}{i}   & 11.2602880(11) & 9.40 & 6.84 &3.61       \\
                \ion{C}{ii}   & 24.383154(16)  & 21.64 &16.55  &        10.13   \\
                \ion{C}{iii}   & 47.88778(25)  &  45.34& 37.63 &27.69           \\
                \ion{C}{iv}   &64.49352(19)   &  62.15& 51.92 &38.80            \\
                \ion{C}{v}   & 392.090518(25) &  390.04& 376.60 &357.05 \\
           \hline
        \end{tabular}
           \tablefoot{\tablefoottext{a}{\citet{nist}.}}
\end{table}

\begin{table}[!ht]
  \caption{Computed ionization potentials for \ion{Si}{i}--\ion{Si}{xiii} as a function of the plasma screening parameter $\mu$ (a.u.). Spectroscopic values (NIST) are also listed for comparison. \label{IPSi}}
  \centering
        \begin{tabular}{lrrrr}
                \hline \hline
                \noalign{\smallskip}
                Ion    & \multicolumn{4}{c}{IP (eV)} \\
                \cline{2-5}
                \noalign{\smallskip}
                &  NIST$^a$ & $\mu = 0.0$ & $\mu = 0.1$ & $\mu = 0.25$ \\
                \hline
                \noalign{\smallskip}            
                        \ion{Si}{i}   & 8.15168(3) & 7.27 & 4.57 &      0.91\\
                        \ion{Si}{ii}   & 16.34585(4)   & 15.11 & 9.98 &3.52     \\
                        \ion{Si}{iii}   & 33.493(9)  & 33.05 & 25.48 & 16.29    \\
                        \ion{Si}{iv}   &45.14179(7)   & 43.62 & 33.63 &21.67    \\
                        \ion{Si}{v}   & 166.767(3) & 158.12 &144.76 &125.77     \\
                        \ion{Si}{vi}   & 205.279(5)  & 202.80 &186.75 &163.89   \\
                        \ion{Si}{vii}   & 246.57(5) & 245.09 &226.39 &199.78    \\
                        \ion{Si}{viii}   & 303.59(5) & 302.35 & 280.98& 250.49\\
                        \ion{Si}{ix}   & 351.28(6)  & 349.44 & 325.40&291.10    \\
                        \ion{Si}{x}   & 401.38(4) & 398.96 & 372.27&334.13      \\
                        \ion{Si}{xi}   &  476.273(19) & 474.22 & 444.93&403.37  \\
                        \ion{Si}{xii}   & 523.415(7) & 521.10 & 489.18&443.93   \\
                        \ion{Si}{xiii}   & 2437.65815(20) & 2437.18 & 2401.89&2349.89\\
           \hline
        \end{tabular}
           \tablefoot{\tablefoottext{a}{\citet{nist}.}}
\end{table}

\begin{table}[!ht]
  \caption{Computed ionization potentials for \ion{Ca}{i}--\ion{Ca}{xix} as a function of the plasma screening parameter $\mu$ (a.u.). Spectroscopic values (NIST) are also listed for comparison. \label{IPCa}}
  \centering
        \begin{tabular}{lrrrr}
                \hline \hline
                \noalign{\smallskip}
                Ion    & \multicolumn{4}{c}{IP (eV)} \\
                \cline{2-5}
                \noalign{\smallskip}
                &  NIST$^a$ & $\mu = 0.0$ & $\mu = 0.1$ & $\mu = 0.25$ \\
                \hline
                \noalign{\smallskip}
                        \ion{Ca}{i}  & 6.11315547(25) & 7.06 & 4.72 &2.46       \\
                        \ion{Ca}{ii}  & 11.871719(4) & 13.38 &  8.60&3.44       \\
                        \ion{Ca}{iii}  & 50.91316(25)   & 44.39 &36.23  &24.46  \\
                        \ion{Ca}{iv}  & 67.2732(21)   & 66.99 & 56.36 &41.67    \\
                        \ion{Ca}{v}  & 84.34(8)   & 86.65 & 73.48 &55.33        \\
                        \ion{Ca}{vi}  & 108.78(25)   & 105.11 &89.32  &67.98    \\
                        \ion{Ca}{vii}  & 127.21(25)   & 123.62 &105.25  &80.23  \\
                        \ion{Ca}{viii}  &147.24(12)   & 143.56 & 122.60 &94.10  \\
                        \ion{Ca}{ix}  &188.54(6)   &189.24  &165.80  &134.40    \\
                        \ion{Ca}{x}  & 211.275(4)  & 210.37 & 184.38 &149.65\\
                        \ion{Ca}{xi}  & 591.60(12)  & 583.58 & 554.00 &511.15   \\
                        \ion{Ca}{xii}  & 658.2(9)  &  656.77&624.52  &577.82    \\
                        \ion{Ca}{xiii}  & 728.6(1.1)  &727.59  &692.67 &642.07         \\
                        \ion{Ca}{xiv}  & 817.2(6)   & 816.38 &778.76  &724.25   \\
                        \ion{Ca}{xv}  & 894.0(4)  & 893.69 &853.40 &794.94      \\
                        \ion{Ca}{xvi}  & 973.7(3)  &972.00  &929.00  &866.64    \\
                        \ion{Ca}{xvii}  & 1086.8(4) &1085.69 &1040.13 &974.32   \\
                        \ion{Ca}{xviii}  & 1157.726(7)  & 1155.72 &1107.48 &1037.88\\
                        \ion{Ca}{xix}  & 5128.8578(5) & 5131.55  &5080.04  &5003.48\\
           \hline
        \end{tabular}
           \tablefoot{\tablefoottext{a}{\citet{nist}.}}
\end{table}

\begin{table}[!ht]
  \caption{Computed ionization potentials for \ion{Cr}{i}--\ion{Cr}{xxiii} as a function of the plasma screening parameter $\mu$ (a.u.). Spectroscopic values (NIST) are also listed for comparison. \label{IPCr}}
  \centering
        \begin{tabular}{lrrrr}
                \hline \hline
                \noalign{\smallskip}
                Ion    & \multicolumn{4}{c}{IP (eV)} \\
                \cline{2-5}
                \noalign{\smallskip}
                &  NIST$^a$ & $\mu = 0.0$ & $\mu = 0.1$ & $\mu = 0.25$ \\
                \hline
                \noalign{\smallskip}
                        \ion{Cr}{i}  & 6.76651(4) & 9.36 & 7.62  &3.13  \\
                        \ion{Cr}{ii}  & 16.486305(15) & 20.6 & 16.48 &8.83      \\
                        \ion{Cr}{iii}  & 30.959(25) & 34.16 &26.75  &15.33      \\
                        \ion{Cr}{iv}  & 49.16(5) &52.93 &42.12  &27.66  \\
                        \ion{Cr}{v}  & 69.46(4) & 65.43 &52.23  &33.20  \\
                        \ion{Cr}{vi}  & 90.6349(7) & 86.71 &70.73  &48.32       \\
                        \ion{Cr}{vii}  & 160.29(6) & 156.73 &138.12  &112.10    \\
                        \ion{Cr}{viii}  & 184.76(15) & 185.46 &164.28  &134.98  \\
                        \ion{Cr}{ix}  & 209.5(3) & 212.53 & 188.78 &155.98      \\
                        \ion{Cr}{x}  & 244.5(5) &  240.10&  213.69&177.11       \\
                        \ion{Cr}{xi}  & 270.8(5) &  266.81& 237.76 &197.53      \\
                        \ion{Cr}{xii}  & 296.7(6) &  294.57&262.91  &219.03     \\
                        \ion{Cr}{xiii}  & 354.7(3) & 355.38 &321.20  &274.28    \\
                        \ion{Cr}{xiv}  & 384.163(6) &  383.55&346.80  &296.37   \\
                        \ion{Cr}{xv}  & 1011.6(5) & 1004.16 &963.74  &904.82    \\
                        \ion{Cr}{xvi}  & 1097.2(1.4) &  1096.29& 1053.19 &990.39 \\
                        \ion{Cr}{xvii}  & 1188.0(2.1) & 1187.57 &1141.80  &1075.06       \\
                        \ion{Cr}{xviii}  & 1294.8(1.6) &  1294.09&1245.62  &1174.95\\
                        \ion{Cr}{xix}  & 1394.5(7) & 1395.68 &1344.52 &1269.85  \\
                        \ion{Cr}{xx}  & 1495.1(7) & 1494.11 & 1440.25 &1361.65  \\
                        \ion{Cr}{xxi}  & 1634.1(5) & 1633.70 & 1577.27 &1495.22 \\
                        \ion{Cr}{xxii}  & 1721.183(7) & 1719.53 & 1660.41 & 1574.54\\
                        \ion{Cr}{xxiii}  & 7481.8628(7)  & 7488.10 & 7425.71 & 7332.83       \\
           \hline
        \end{tabular}
           \tablefoot{\tablefoottext{a}{\citet{nist}.}}
\end{table}

\begin{table}[!ht]
  \caption{Computed ionization potentials for \ion{Ni}{i}--\ion{Ni}{xxvii} as a function of the plasma screening parameter $\mu$ (a.u.). Spectroscopic values (NIST) are also listed for comparison. \label{IPNi}}
  \centering
        \begin{tabular}{lrrrr}
                \hline \hline
                \noalign{\smallskip}
                Ion    & \multicolumn{4}{c}{IP (eV)} \\
                \cline{2-5}
                \noalign{\smallskip}
                &  NIST$^a$ & $\mu = 0.0$ & $\mu = 0.1$ & $\mu = 0.25$ \\
                \hline
                \noalign{\smallskip}
                        \ion{Ni}{i}  &  7.639878(17)  & 9.29  & 6.99 & 3.56\\
                        \ion{Ni}{ii}  & 18.168838(25) & 20.92 & 15.45 &8.75     \\
                        \ion{Ni}{iii}  & 35.187(19) & 31.86 & 23.78 & 13.66\\
                        \ion{Ni}{iv}  & 54.92(25) & 59.20 & 48.49 &     34.91\\
                        \ion{Ni}{v}  & 76.06(6) & 73.19 & 60.07 &42.32\\
                        \ion{Ni}{vi}  & 108(1) & 117.94 & 102.11 &77.83 \\
                        \ion{Ni}{vii}  & 132(2) & 122.09 & 103.20 & 80.05       \\
                        \ion{Ni}{viii}  & 162.0(2.1) & 167.48 &146.22  &        116.63\\
                        \ion{Ni}{ix}  & 193.2(5)  & 190.33 & 166.30 &132.35     \\
                        \ion{Ni}{x}  & 224.7(5) &  221.07& 194.41 &     156.74\\
                        \ion{Ni}{xi}  & 319.5(7) & 317.29 & 288.03 &246.89      \\
                        \ion{Ni}{xii}  & 351.6(3) &  353.13&321.28  &276.68     \\
                        \ion{Ni}{xiii}  & 384.5(5) &  388.19&353.72  &305.50    \\
                        \ion{Ni}{xiv}  & 429.3(8) & 423.63 &386.49  &334.40     \\
                        \ion{Ni}{xv}  & 462.8(1.1) &  459.54&  419.75&363.90    \\
                        \ion{Ni}{xvi}  & 495.4(1.7) &495.37  &452.94  &393.35   \\
                        \ion{Ni}{xvii}  & 571.07(12) & 571.44 &526.47  &463.75  \\
                        \ion{Ni}{xviii}  & 607.000(19) & 606.71 &559.14  &492.81 \\
                        \ion{Ni}{xix}  & 1541.0(8) &  1534.10& 1482.83 &1407.77 \\
                        \ion{Ni}{xx}  & 1646(3) & 1645.27 &1591.32  &1512.34    \\
                        \ion{Ni}{xxi}  & 1758(4) &1758.28  &1701.64  &1618.71   \\
                        \ion{Ni}{xxii}  & 1880(5) & 1880.06 &1820.73  &1733.83  \\
                        \ion{Ni}{xxiii}  & 2008.1(1.3) &2011.26  &1949.22  &1858.31       \\
                        \ion{Ni}{xxiv}  & 2130.5(9) & 2130.33 & 2065.61 &1970.74\\
                        \ion{Ni}{xxv}  & 2295.6(2.1) &2296.14  &2228.82  &2130.51        \\
                        \ion{Ni}{xxvi}  & 2399.259(7) &2398.12 & 2328.14 &2225.96        \\
                        \ion{Ni}{xxvii}  & 10288.8862(14) & 10300.17  &10226.89  &10117.68      \\      
           \hline
        \end{tabular}
           \tablefoot{\tablefoottext{a}{\citet{nist}.}}
\end{table}

\pagebreak

\newpage

\section{Computed K-threshold energies for different plasma screening parameter values} 

\begin{table}[!ht] 
  \caption{Computed K-thresholds for \ion{C}{i}--\ion{C}{v} as a function of the plasma screening parameter $\mu$ (a.u.). \label{KTC}}
  \centering
        \begin{tabular}{lrrr}
                \hline \hline
                \noalign{\smallskip}
                Ion & \multicolumn{3}{c}{$E_K$ (eV)} \\
                \cline{2-4}
                \noalign{\smallskip}
                & $\mu = 0.0$       & $\mu = 0.1 $ & $\mu = 0.25$ \\
                \hline
                \noalign{\smallskip}
                        \ion{C}{i}   & 289.75 &287.04 &282.90   \\
                        \ion{C}{ii}   & 309.54 &304.15 &296.20          \\
                        \ion{C}{iii}  &  334.77&326.65 &314.67  \\
                        \ion{C}{iv}  & 359.45 &348.68 &332.95           \\
                        \ion{C}{v}  & 390.04 &376.60 &357.05            \\
                \hline
        \end{tabular}
\end{table}

\begin{table}[!ht] 
  \caption{Computed K-thresholds for \ion{Si}{i}--\ion{Si}{xiii} as a function of the plasma screening parameter $\mu$ (a.u.). \label{KTSi}}
  \centering
        \begin{tabular}{lrrr}
                \hline \hline
                \noalign{\smallskip}
                Ion & \multicolumn{3}{c}{$E_K$ (eV)} \\
                \cline{2-4}
                \noalign{\smallskip}
                & $\mu = 0.0$       & $\mu = 0.1 $ & $\mu = 0.25$ \\
                \hline
                \noalign{\smallskip}                    
                        \ion{Si}{i}   &1821.26  &1817.92 &1813.59       \\
                        \ion{Si}{ii}   & 1838.08 &1832.20 & 1821.90     \\
                        \ion{Si}{iii}   &1858.17  & 1849.79& 1836.43    \\
                        \ion{Si}{iv}  & 1874.94 &1864.04 & 1847.64      \\
                        \ion{Si}{v}   &  1894.25&1880.62 &1860.04       \\
                        \ion{Si}{vi}   &  1953.25&1936.89 &1912.31      \\
                        \ion{Si}{vii}   &2010.49  &1991.46 &1963.00     \\
                        \ion{Si}{viii}   &2070.35  &2048.64 &2016.25    \\
                        \ion{Si}{ix}   &  2140.72&2115.91 &2079.60      \\
                        \ion{Si}{x}   &  2214.64&2187.54 &2147.29       \\
                        \ion{Si}{xii}   &  2360.41&2327.90 &2279.73     \\
                        \ion{Si}{xiii}   &  2437.18&2401.89 &2349.89    \\
                \hline
        \end{tabular}
\end{table}

\begin{table}[!ht] 
  \caption{Computed K-thresholds for \ion{Ca}{i}--\ion{Ca}{xix} as a function of the plasma screening parameter $\mu$ (a.u.). \label{KTCa}}
  \centering
        \begin{tabular}{lrrr}
                \hline \hline
                \noalign{\smallskip}
                Ion & \multicolumn{3}{c}{$E_K$ (eV)} \\
                \cline{2-4}
                \noalign{\smallskip}
                & $\mu = 0.0$       & $\mu = 0.1 $ & $\mu = 0.25$ \\
                \hline
                \noalign{\smallskip}                    
                        \ion{Ca}{i}  &4025.91 &4022.24 &4009.91\\
                        \ion{Ca}{ii}  & 4031.96 &4025.96 &4011.17       \\
                        \ion{Ca}{iii}   & 4041.12 &4031.87 &4014.35     \\
                        \ion{Ca}{iv}   & 4069.19 &4057.48 &4037.11      \\
                        \ion{Ca}{v}   & 4094.19 &4079.97 &4056.54       \\
                        \ion{Ca}{vi}   & 4116.98 &4100.25 &4073.38\\
                        \ion{Ca}{vii}   & 4148.39 &4128.98 &4098.60     \\
                        \ion{Ca}{viii}   & 4182.66 &4160.66 &4126.84    \\
                        \ion{Ca}{ix}   & 4220.38 &4195.80 &4158.58      \\
                        \ion{Ca}{x}   & 4252.66 &4225.46 &4184.74       \\
                        \ion{Ca}{xi}   & 4288.88 &4258.97 &4214.21      \\
                        \ion{Ca}{xii}   &  4386.01& 4353.41&4304.73     \\
                        \ion{Ca}{xiii}   & 4480.87 &4445.57 &4392.93    \\
                        \ion{Ca}{xiv}   & 4578.38 &4540.38 &4483.77     \\
                        \ion{Ca}{xv}   &  4688.52&4647.82 &4587.21      \\
                        \ion{Ca}{xvi}   & 4802.13 &4758.71 &4694.13     \\
                        \ion{Ca}{xvii}   & 4921.17 &4875.07 &4806.49    \\
                        \ion{Ca}{xviii}   &5019.81  &4970.98 &4898.49   \\
                        \ion{Ca}{xix}   &5131.55  &5080.04  &5003.48    \\      
                \hline
        \end{tabular}
\end{table}

\begin{table}[!ht] 
  \caption{Computed K-thresholds for \ion{Cr}{i}--\ion{Cr}{xxiii} as a function of the plasma screening parameter $\mu$ (a.u.). \label{KTCr}}
  \centering
        \begin{tabular}{lrrr}
                \hline \hline
                \noalign{\smallskip}
                Ion & \multicolumn{3}{c}{$E_K$ (eV)} \\
                \cline{2-4}
                \noalign{\smallskip}
                & $\mu = 0.0$       & $\mu = 0.1 $ & $\mu = 0.25$ \\
                \hline
                \noalign{\smallskip}
                        \ion{Cr}{i}   & 5986.24 &5983.20 &5971.52       \\
                        \ion{Cr}{ii}   &  5990.75& 5984.40&5969.91      \\
                        \ion{Cr}{iii}   & 5998.42 &5989.74 &5971.51     \\
                        \ion{Cr}{iv}   & 6027.68 &6015.91 &5995.02      \\
                        \ion{Cr}{v}   &  6045.78& 6031.43&6006.19       \\
                        \ion{Cr}{vi}   & 6074.89 &6057.75 &6029.14      \\
                        \ion{Cr}{vii}   & 6106.54 &6086.76 &6054.57     \\
                        \ion{Cr}{viii}   & 6144.53 &6122.20 &   6086.79\\
                        \ion{Cr}{ix}   & 6180.81 &6155.91 &6117.14      \\
                        \ion{Cr}{x}   & 6215.57 &6188.01 &6145.49       \\
                        \ion{Cr}{xi}   & 6258.95 &6228.77 &6182.67      \\
                        \ion{Cr}{xii}   & 6305.00 &6272.21 &6222.51     \\
                        \ion{Cr}{xiii}   & 6354.70 &6319.28 &6266.03    \\
                        \ion{Cr}{xiv}   & 6397.39 &6359.34 &6302.44     \\
                        \ion{Cr}{xv}   & 6444.78 &6404.00 &6343.07      \\
                        \ion{Cr}{xvi}   & 6566.86 &6523.40 &6458.53     \\
                        \ion{Cr}{xvii}   & 6686.93 &6640.77 &6571.93    \\
                        \ion{Cr}{xviii}   & 6809.69 &6760.82 &6687.98   \\
                        \ion{Cr}{xix}   & 6947.12 &6895.54 &6818.68     \\
                        \ion{Cr}{xx}   & 7086.53 &7032.24 &6951.38      \\
                        \ion{Cr}{xxi}   & 7232.77 &7175.79 &7090.93     \\
                        \ion{Cr}{xxii}   & 7353.27 &7293.00 &7204.11    \\
                        \ion{Cr}{xxiii}   & 7488.10 &7425.71 &7332.83   \\
                \hline
        \end{tabular}
\end{table}

\begin{table}[!ht] 
  \caption{Computed K-thresholds for \ion{Ni}{i}--\ion{Ni}{xxvii} as a function of the plasma screening parameter $\mu$ (a.u.). \label{KTNi}}
  \centering
        \begin{tabular}{lrrr}
                \hline \hline
                \noalign{\smallskip}
                Ion & \multicolumn{3}{c}{$E_K$ (eV)} \\
                \cline{2-4}
                \noalign{\smallskip}
                & $\mu = 0.0$       & $\mu = 0.1 $ & $\mu = 0.25$ \\
                \hline
                \noalign{\smallskip}
                        \ion{Ni}{i}   & 8338.93 & 8335.97 & 8329.38\\
                        \ion{Ni}{ii}   &  8352.17&8346.55 &8336.12      \\
                        \ion{Ni}{iii}   & 8366.25 &8357.58 &8342.02     \\
                        \ion{Ni}{iv}   &  8385.24&8373.79 &8354.62      \\
                        \ion{Ni}{v}   &  8394.95& 8380.81&8357.85       \\
                        \ion{Ni}{vi}   &  8442.32&8425.37 &8398.20      \\
                        \ion{Ni}{vii}   & 8460.52 &8440.43 &8409.02     \\
                        \ion{Ni}{viii}   & 8507.20 &8484.81 &8448.98    \\
                        \ion{Ni}{ix}   &  8538.17&8513.01 &8472.91      \\
                        \ion{Ni}{x}   & 8579.88 &8552.08 &8508.25       \\
                        \ion{Ni}{xi}   & 8624.03 &8593.57 &8546.01      \\
                        \ion{Ni}{xii}   & 8673.09 &8640.03 &8589.06     \\
                        \ion{Ni}{xiii}   & 8720.91 & 8685.25&8630.76    \\
                        \ion{Ni}{xiv}   & 8767.56 &8729.22 &8670.97     \\
                        \ion{Ni}{xv}   & 8823.07 & 8782.09&8720.06      \\
                        \ion{Ni}{xvi}   & 8880.92 &8837.29 &8771.57     \\
                        \ion{Ni}{xvii}   & 8942.78 & 8896.51&8827.13    \\
                        \ion{Ni}{xviii}   & 9002.30 &8953.40 &8880.34   \\
                        \ion{Ni}{xix}   & 9054.58 &9002.95 & 8925.80    \\
                        \ion{Ni}{xx}   & 9201.61 &9147.28 &9066.18      \\
                        \ion{Ni}{xxi}   & 9347.19 &9290.16 &9205.06     \\
                        \ion{Ni}{xxii}   & 9495.75 &9436.01 &9346.90    \\
                        \ion{Ni}{xxiii}   & 9660.58& 9598.12&9504.99    \\
                        \ion{Ni}{xxiv}   & 9825.48 &9760.33 &9663.18    \\
                        \ion{Ni}{xxv}   & 9999.35 & 9931.48&9830.32     \\
                        \ion{Ni}{xxvi}  & 10140.77 & 10070.20 &9964.99  \\
                        \ion{Ni}{xxvii} & 10300.17  &10226.89  &10117.68 \\              
                \hline
        \end{tabular}
\end{table}

\end{appendix}

\end{document}